\documentclass[twocolumn]{openjournal} 

\usepackage[T1]{fontenc}
\usepackage{ae,aecompl}

\usepackage{graphicx}	% Including figures
\usepackage{amsmath}	% Advanced maths commands
\usepackage{amssymb}	% Extra maths symbols
\usepackage{fancyvrb}
\usepackage{etoolbox}
\makeatletter
\patchcmd{\@verbatim}
{\verbatim@font}
{\verbatim@font}
{}{}
\makeatother
\usepackage{natbib}

\usepackage{hyperref}
\hypersetup{
	colorlinks=true,
	urlcolor=blue,    % color of external links
	linkcolor=black,  % color of toc, list of figs etc.
	citecolor=blue,   % color of links to bibliography
}

\defcitealias{Pittordis_2018}{PS18}
\defcitealias{Pittordis_2019}{PS19}
\defcitealias{Pittordis_2023}{PS23}
\defcitealias{Badry_2021}{ERH}
\defcitealias{Banik_2018_Centauri}{BZ18}

\renewcommand\today{2025 July 30  }   % Kill the horrible confusing  \today command.   Use \draftdate below. 

\newcommand{\draftdate}{ Accepted by Open Journal of Astrophysics - 2025 July 29  } 
\newcommand{\kms} {{\rm \, km \, s^{-1} }} 
\newcommand{\GAIA} {{\em Gaia\ }}
\newcommand{\au} {\, {\rm AU}}   
\newcommand{\kau} {\, {\rm kAU}} 
\newcommand{\msun} {\,M_\odot} 
 
\newcommand{\pc} {\, {\rm pc}} 
 
\newcommand{\msecsq}{\, {\rm m \, s^{-2}}}

\newcommand{\dvp}{\Delta v_{\rm p} } % new default is Delta v_p  
\newcommand{\dmg}{\Delta M_{G} }   % for lobster 
\newcommand{\dmgone}{\Delta M_{G, 1} }   % for lobster 
\newcommand{\dmgtwo}{\Delta M_{G, 2} }   % for lobster 
 
\newcommand{\vtilde}{\tilde{v}} 
 
\newcommand{\rp} {r_{\rm p}} 
\newcommand{\ftrip}{f_{\rm trip}} 
\newcommand{\mas} {\, {\rm mas}} 
 
\newcommand{\tent}[1] {\times 10^{#1} }    

 \newcommand{\newa}{  }  

\begin{document}
	
	\title[Wide Binaries, triples, GR vs MOND]   % short title in square brackets 
  {Wide Binaries from Gaia DR3 : testing GR vs MOND  with realistic triple modelling}
	 \date{\draftdate}
	
	\author{C. Pittordis} 
	\email[E-mail: ]{cp.pittordis@gmail.com}
	\author{W. Sutherland} 
	\email[E-mail: ]{w.j.sutherland@qmul.ac.uk}
          \author{P. Shepherd} 
          \email[E-mail: ]{p.d.shepherd@qmul.ac.uk} 
	\affiliation{School of Physical and Chemical Sciences,  
		Queen Mary University of London, Mile End Road, London E1 4NS, UK.}

	\begin{abstract} 
		We provide an updated test for modifications of gravity from a sample of wide-binary stars from \GAIA DR3, and their sky-projected relative velocities.  Here we extend on our earlier 2023 study, using several updated selection cuts aimed at reducing contamination from triple systems with an undetected third star. We also use improved mass estimates from \GAIA FLAME, and we add refinements to previous modelling of the triple and other populations and 
 the model-fitting.  We fit histograms of observed vs Newtonian velocity differences 
 to a flexible mixture of binary + triple populations with realistic eccentricity distributions, plus unbound flyby and random-chance populations.   
We find as before that Newtonian models provide a significantly better fit than MOND, though improved   understanding of the triple population is necessary to make this fully decisive. .   
		
	\end{abstract}
	\maketitle
	
	\section{Introduction}
	A number of recent studies have analysed relative velocities of very wide binary stars in \GAIA DR3 to attempt to discriminate between GR + dark matter models and modified-gravity MOND-like models (see \citet{Famaey_McGaugh_2012} for a review of MOND models).  This wide-binary test is potentially 
powerful, since dark-matter models predict that wide-binary stars should contain negligible dark matter
	and thus be entirely Newtonian,  whereas MOND-like models predict increased accelerations leading to faster relative velocities  at scales where the 
	internal accelerations are below the MOND characteristic acceleration scale $a_0 \simeq 1.2 \times 10^{-10} \msecsq$. 
	This acceleration occurs at separation $\simeq 7 \kau$ for a binary of $1 \msun$ total mass, while there is a smooth transition (with the details dependent on the specific version of MOND),  so MOND predicts some deviations should be observable down to smaller separations $\sim 2 - 4 \kau$.   
	The first studies of this wide-binary test were by \citet{Hernandez_2012}; \citet{Hernandez_2012_2}; \citet{Hernandez_2014}, but these early studies
	were hampered by the limited precision of ground-based proper motion data.  
	In our first paper in this series \citep[][hereafter \citetalias{Pittordis_2018}]{Pittordis_2018}
	we used simulations to highlight the dramatic improvement then
	anticipated from the \GAIA spacecraft \citep{Prusti_2016}.  This was followed by our selection of wide binaries observed in \GAIA DR2 
	\citep[][hereafter \citetalias{Pittordis_2019}]{Pittordis_2019}: 
	that study revealed a ``long tail" of apparent binaries with relative velocities well above expectations from either GR or MOND, 
	and \citep{Clarke_2020} suggested a likely explanation was triple systems with an unresolved or unseen third star perturbing the velocity of its close companion; this induces an additional term which is the relative velocity
	between the photocentre and the barycentre of the close binary, which often dominates over the 
	relative velocity between the distant star and the barycentre of the close binary.  
 The model of formation of wide binaries by \citet{Kouwenhoven_2010} provides some support to this, 
  predicting that a fairly substantial fraction of observed wide ``binary" systems should actually be triples or quadruples   with close inner system(s).

	More recently, a number of wide-binary studies have been done using \GAIA EDR3 and DR3, with contrasting results: 
	\citet{Hernandez_2022} and \citet{Hernandez_2023} analysed relatively small samples with stringent data-quality cuts and claim evidence for a MOND signal.  Also \citet{Chae_2023} and \citet{Chae_2024} have selected larger samples and also claim evidence for a MOND signal.  Conversely, 
\citet[][hereafter \citetalias{Pittordis_2023}]{Pittordis_2023}
  fitted a mixed model 
	of binaries, triples and flyby systems to their own sample of wide binary candidates, and found
	a significant preference for GR over MOND. Also \citet{Banik_2024} used a subset of the PS23 binary sample, 
	and used a sophisticated likelihood-modelling based on a large library of binary and triple orbits, and
	also found that GR was preferred over MOND at high significance. 
	
	These apparently contradictory results appear to originate from differences in sample selection, 
	statistical methods and/or modelling of triple systems, with the triple systems perhaps the dominant 
	source of uncertainty. 
	
	Fortunately for future prospects, a study by \citet{Manchanda_2023} showed 
  that a high fraction $\sim 85\%$ to $95\%$ 
	of unseen triple systems
	can be flagged by combining a variety of followup methods, including future \GAIA multi-epoch astrometry, and followup speckle-imaging and coronagraphic observations; a key conclusion from \citet{Manchanda_2023}  
   is that nearly all main-sequence third stars are detectable in principle: third stars closer than $\la 25 \au$ can largely be detected by astrometric accelerations in \GAIA data, 
	while third stars at separations $\ga 20 \au$ are detectable by speckle and/or coronagraphic imaging.  
	However, this scenario requires both the full 10-year mission \GAIA data (DR5 projected for 2030) and 
	substantial telescope time for followup imaging, so is a long-term prospect. In the short term it is desirable
	to revisit the gravity test based on the available \GAIA DR3 data, which is the purpose of this paper. 
	
	This paper is essentially an updated version of \citet{Pittordis_2023} with several improvements aimed
	mainly at reducing the fraction of triple systems via data cuts, more realistic modelling of the triple
 population, and an improved fitting procedure; 
    The main refinements are the following: 
 \begin{itemize} 
\item[i)]  A tighter cut on the {\tt ruwe} parameter, 
\item[ii)] A new cut against triple systems using the ``Lobster" diagram, 
\item[iii)] Improved stellar mass estimates using the \GAIA FLAME data,
\item[iv)] Several enhancements to the model triple distribution. 
\item[v)]  Analysing a wider range of projected separations extending down into the quasi-Newtonian regime,
\item[vi)] An improved fitting procedure with a variable triple fraction, while distributions  
   of unbound systems are constrained to have a realistic distribution of projected separation. 
\end{itemize} 	

	The plan of the paper is as follows: in Section~\ref{sec:dr3} we describe the sample selection 
	and the refinements compared to PS23. In Section~\ref{sec:orbits} we review some statistical properties
	of the $\vtilde$ parameter used later, and describe the modelling of triple and other populations; and in   Section~\ref{sec:fits} we fit the observed $\vtilde$
	histograms as a mixture of binary, triple and flyby populations from either GR or MOND 
	orbit simulations. We summarise our conclusions in Section~\ref{sec:conc}. 
	
%	\newpage   % dont need this ... 	
	\section{ \GAIA DR3 binary sample selection and FLAME masses} 
	\label{sec:dr3} 
	
	\subsection{Preliminary selection} 
	\label{sec:prelim}
	Our starting point is the public \GAIA Data Release 3 dataset (DR3),  
	\citep{GAIA_DR3_2022, GAIA_DR3_2023} released on 2022 June 13, joined with the  Final Luminosity Age Mass Estimator (FLAME) data \citep{GAIA_DR3_Doc_2022}. 
	We initially select all stars with measured parallax $\omega > \frac{10}{3} \mas$
	(i.e. estimated distance $< 300 \pc$) with a \GAIA broadband magnitude $G < 17$, and cutting out the Galactic plane with absolute latitude $\vert b \vert \le 15 \deg$, yielding a preliminary DR3 sample of 2,102,657 stars (hereafter PDR3).  The ADQL query used is given in the Appendix.  Star data-quality cuts are applied at a later stage, in order that these may be adjusted after the initial selection.  
	
	We then applied a similar search method as described in section 2 from \citetalias{Pittordis_2019} and \citetalias{Pittordis_2023} to PDR3, to search this nearby-star sample for pairs of stars
	with projected separation $\le 50 \kau$ (calculated 
	at the mean distance of each candidate pair), and distances  
	consistent with each other within $4\times$
	the combined uncertainty (with an upper limit of $8 \pc$) i.e.  
    $ \vert d_1 - d_2 \vert \leq min( 4\sigma_{d} , 8 \pc )$,  and projected velocity difference $\dvp \le 5 \kms$ 
	as inferred from the difference in proper motions; here, the projected velocity difference is computed  
	assuming {\em both} stars in each candidate pair are
	actually at the mean of the two estimated distances.  (Note here, this $5 \kms$ velocity difference is 
    enlarged from the $3 \kms$ used by \citetalias{Pittordis_2023},  to avoid incompleteness for the
 high-velocity tail of systems at smaller  projected separations, down to $1.25 \kau$ analysed below). 
		
	This search results in a first-cut
	sample of 97,505  candidate DR3 wide binaries, including FLAME masses where available 
 (hereafter WB-DR3-FLAME).  
% {\bf Reword because missing FLAME masses}. \\ 

\subsection{Additional cuts} 
\label{sec:add-cuts} 
 This sample is then pruned using the same additional cuts as section 2.2 - 2.5 of \citetalias{Pittordis_2023},  for removing moving groups; known open clusters; systems with a fainter nearby companion at $G < 20$ and
 parallax consistent with equal distance;  and a cut on 
 the $u$ parameter in Eq.~1 of \citetalias{Pittordis_2023}.  This reduces the WB-DR3-FLAME sample to 75,501
 candidates.  
	
	Additional cuts are applied to this sample based on the ``Renormalised Unit Weight Error" or {\tt ruwe}  parameter defined in \GAIA DR3: this is a measure of scatter of single-epoch \GAIA observations around the 
	basic 5-parameter fit parallax + uniform proper motion, rescaled by a factor
   dependent on magnitude and colour so the median {\tt ruwe} for single stars is close to 1.  Objects
with a {\tt ruwe} value significantly larger than 1 are indicative of excess scatter which may indicate a poor fit 
or astrometric wobble from an unresolved close binary. The studies by \cite{Belokurov_2020} and \citet{Fitton_2022} analysed the \GAIA {\tt ruwe} value for single stars, and show an increased probability of binarity at {\tt ruwe} $ >  1.2$; therefore we apply an additional cut that both stars in a candidate binary are required to have {\tt ruwe} $ < 1.2$:  note that this is more restrictive than the {\tt ruwe} $< 1.4$ used in \citetalias{Pittordis_2023}.  
	 
	In addition, we also apply a cut based on the ``Image Parameters Determination of Multiple Peaks'', the parameter  {\tt ipd$\_$frac$\_$multi$\_$peak} defined in \GAIA DR3;
	this parameter provides information on the raw windows used for the astrometric processing of this source from the Image Parameters Determination (IPD) module in the core processing. It is defined as the integer percentage of windows (having a successful IPD result), for which the IPD algorithm has identified a double peak, meaning that the detection may be a visually resolved double star (either just visual double or real binary). 
The study by \citet{Tokovinin_2023}, comparing nearby hierarchical systems with \GAIA and speckle interferometry, indicates that resolved pairs have values {\tt ipd\_frac\_multi\_peak}~$>2$.  Therefore, we apply a cut that both stars in our candidate wide binary have {\tt ipd\_frac\_multi\_peak }~$\leq 2$.  (Small positive values of 1 or 2 may occur
 for single stars from an occasional cosmic-ray hit etc, and are accepted). 
	
After both cuts {\tt ruwe} $ < 1.2$ and {\tt ipd$\_$frac$\_$multi$\_$peak} $ \leq 2$, our WB-DR3-FLAME sample is reduced to 40,116 candidate wide binaries.  
\vfill 

	\subsection{Mass Estimates using FLAME}
	\label{sec:mass-estimate-flames}
In the preliminary selection \ref{sec:DR3_FLAMES_QUERY} creating the PDR3, only $\sim 35$\% of the stars have an estimated  FLAME mass.  In our WB-DR3-FLAME binaries, 
 only 17\% have a FLAME mass for both stars, while 62\% have a FLAME mass for at least one star. For our gravity test below, we require a mass estimate for each binary candidate, hence we estimate masses for the stars lacking
 FLAME masses via a combination of the sample from \citet{Pecaut_2013} and the populated FLAME mass candidates. We begin by adopting the main-sequence $M_I(mass)$ relation of \textit{Version 2021.03.02} from \citet{Pecaut_2013} for the mass range $0.18 \leq M/ M_{\odot} \leq 2.0 $, 
	and the $V - I, M_I$ colour relation from the same,
	where $M_I$ denotes I-band absolute magnitude.  We then apply the colour relation given in Table~C2 (i.e., Johnson-Cousins relation) of \citet{GaiaEDR3_2021} to predict
	$G$ magnitude from $V$ and $I$ magnitudes as 
	\begin{eqnarray} 
		G & \simeq & V \; - \; 0.01597 \; + \; 0.02809 (V-I) \; 
		- \; 0.2483 (V-I)^2 \nonumber \\ 
		& & + 0.03656 (V-I)^3 \; - \; 0.002939 (V-I)^4  
		\label{eq:gvi} 
	\end{eqnarray} 
	to obtain a sub-dataset (hereafter; \textit{PM-GMag-Mass}) of absolute \GAIA magnitude, $M_G$, and mass, in  mass range  $0.18 \leq M /  M_{\odot} \leq 2.0 $ as above. 
	Next, for each star in WB-DR3-FLAME we compute $M_G$ directly from $G$ and 
	parallax distance, then generate a sub-dataset (hereafter; \textit{PS-DR3-FLAME-GMag-Mass}) of \GAIA magnitude $M_G$ and FLAME masses. This follows from joining the two sub-datasets \textit{PM-GMag-Mass} and \textit{PS-DR3-FLAME-GMag-Mass} into one.   
 From this sample, we fit the mass/$M_G$ relation for the
 mass range as above  with a 6th order polynomial,  
		\begin{equation}
		\log_{10} (M / \msun )  =  \sum_{0}^{6} b_n \, M_{G}^n  
		\label{eq:poly-mass-6-order} 
	\end{equation}
giving coefficients $b_0 \simeq 0.505$, $b_1 \simeq -0.125$, $b_2 \simeq -0.0140$, $b_3 \simeq 6.97 \tent{-3}$, $b_4 \simeq -9.33 \tent{-4}$, $b_5 \simeq 5.53\tent{-5}$, $b_6 \simeq -1.38\tent{-6}$.   We then use equation \ref{eq:poly-mass-6-order} to estimate the masses of the stars without FLAME mass estimates 
 from their $M_G$  absolute magnitudes. 
	
\subsection{Lobster diagram cuts}
\label{sec:lobster} 
Here, we apply a further cut to our WB-DR3-FLAME sample using the \textit{Lobster Diagram} technique. This follows the method introduced by \citet{Hartman_2022}, where they cross-matched \& examined \GAIA, SUPERWIDE, TESS, K2 and Kepler data, producing a technique to distinguish between ``pure" wide binaries 
 with exactly 2 stars, and triples where one component of the WB is an unresolved close binary: 
  The method involves computing for each star the
 difference between observed absolute magnitude, and that predicted from a main-sequence colour-magnitude
 relation:  each wide-binary then produces a point on a 2D plot, called the ``Lobster diagram" by \citet{Hartman_2022}.   
 Since deviations from the main-sequence ridgeline due to age and metalliicity variation are highly correlated for both stars in a WB, pure 2-star WBs populate a thin diagonal stripe at 45 degrees (the lobster ``body"); while a luminous third star produces deviations off this line by up to 0.75 mag. These triple systems populate a pair of ``lobster claws" in the plot (see Fig~7 of \citealt{Hartman_2022} ): which claw is occupied depends  
  on which one of the wide pair  is the close binary. 

We begin by creating our sub-dataset from \citet{Pecaut_2013} for the mass range $0.18 \leq M_{\odot} \leq 2.0 $ to include colours $B_p - R_p \leq 3.0$ i.e., for F, G, K, M Stars, and compute the absolute Gaia magnitude $M_G$ magnitude from $V$ and $I$, using the Johnson-Cousins relation \ref{eq:gvi}. Next, we fit a 5th order polynomial to the $M_G$/colour relation as 
\begin{equation}
	M_{G,poly} = \sum_{0}^{5} c_n \, u^n
	\label{eq:poly-mass-5-order} 
\end{equation}
where $u \equiv B_P - R_P - 1.4$, defined so the range in $u$ is roughly symmetric around zero.  The resulting
 fit coefficients are $c_0 \simeq 6.997$, $c_1 \simeq 3.229$, $c_2 \simeq -1.282$, $c_3 \simeq 0.608$, $c_4 \simeq 0.473$, $c_5 \simeq -0.269$.

\begin{figure} 
		\begin{center} 
	\includegraphics[width=\linewidth]{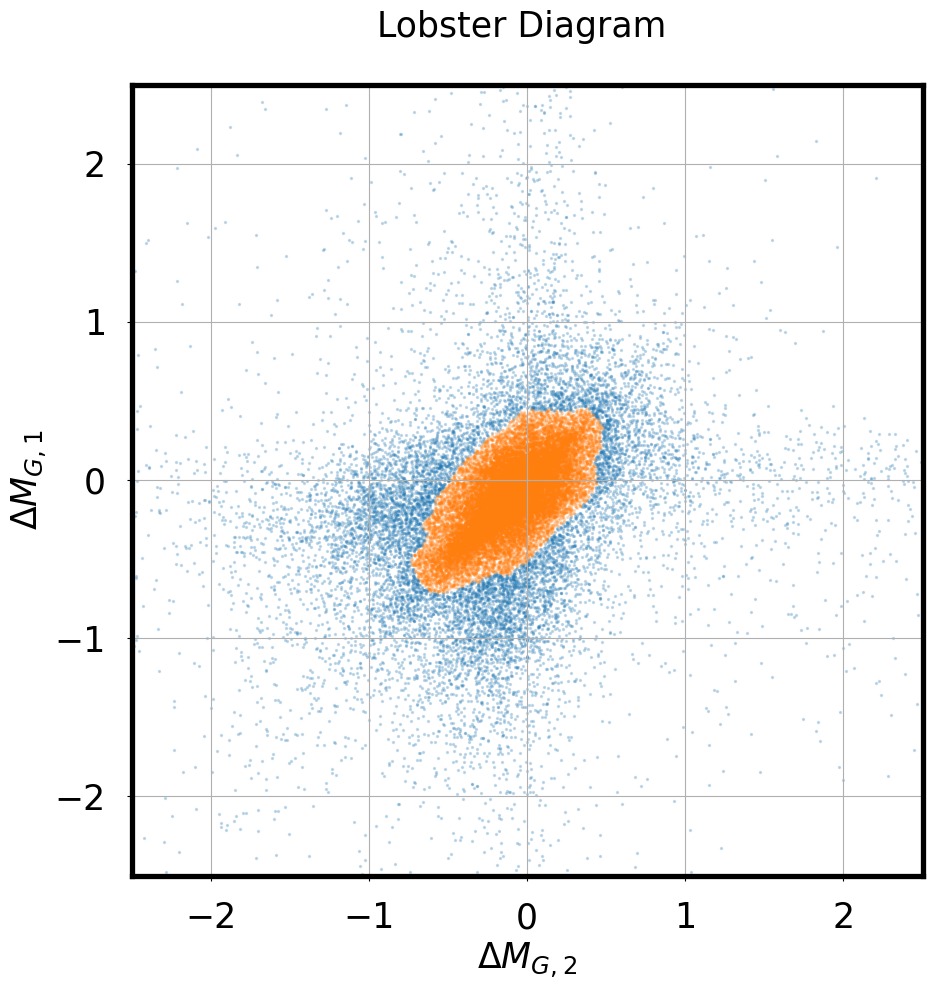} 
			\caption{ 
				Scatter plot comparing the $\dmg$ parameter of Eq.~\ref{eq:lobster-def} for Star 1 \& Star 2 (ordered by ascending declination) of each wide binary system. The panel shows the full 40,116 candidates prior to
 the Lobster cut; the orange points are the subsample of 23,223 candidates selected by 
 Eq.~\ref{eq:model-lobster-body}  } 
			\label{fig:lobster-vetting} 
		\end{center}
	\end{figure}

Given this polynomial, we select binaries from the WB-DR3-FLAME sample where both stars 
 are in the colour range $0.068 \le B_P - R_P \le 3.16$,  
then construct a ``lobster" diagram following \citet{Hartman_2022}: for each star
 in each candidate binary, we define $\Delta M_G$ as the difference between observed $M_G$
 and the fit-line at the observed colour, i.e. 
 \begin{equation} 
  \dmg  \equiv M_{G,obs} - M_{G,poly} 
\label{eq:lobster-def} 
\end{equation} 
Plotting $ \dmgone $ vs $\dmgtwo$ for both stars in each binary system gives a lobster plot shown in Figure~\ref{fig:lobster-vetting}. 
%The reason for this is that scatter around the $M_G / colour$ relation is  
% dominated by age and metallicity variations; for a pure wide binary these are nearly identical
% so these systems tend to lie near the 45-degree line, forming the lobster ``body".  If one component of a wide %binary  is itself an unresolved close binary of comparable luminosities, the point moves systematically away from %the 45-degree line by up to 0.75 mag,  and these systems populate a pair of "lobster claws", 
 %whch claw depending whether star 1 or 2 is the close binary. 

Next, we take values of the two over-luminosity columns $\dmgone , \dmgtwo$ and create a matrix array, and feed this into a Density-Based Spatial Clustering of Applications with Noise (DBSCAN) machine-learning algorithm, where this algorithm is accessed via python package scikit-learn \cite{scikit-learn}, where the DBSCAN algorithm is developed by \cite{DBSCAN_1996} and \cite{DBSCAN_2017}. DBSCAN finds core samples of high density and expands clusters from them. This is ideal for datasets that contain clusters of similar density, in our case, trying to separate the \textit{lobster-body} cluster from the WB-DR3-FLAME dataset itself. After inspecting multiple parameters for our DBSCAN model and analysing their results, the parameters that provided a good model for the lobster-body from our sample are;
\begin{eqnarray}
	Lobster Body  =  DBSCAN( \nonumber \\
		    eps=0.08, \nonumber \\
		    min\_samples=160, \nonumber \\
			 metric=manhattan)
	\label{eq:model-lobster-body} 
\end{eqnarray}
Where {\tt eps} is the maximum distance between two samples 
 (e.g., values from $\dmgone$ and $\dmgtwo$) for one to be considered as in the neighbourhood of the other, $min\_samples$ is the number of samples in a neighbourhood for a point to be considered as a core point. This includes the point itself. If $min\_samples$ is set to a higher value, DBSCAN will find denser clusters, whereas if it is set to a lower value, the found clusters will be more sparse. The $metric$ is the method used to calculate the distances between points, we chose $'manhattan'$ over $`euclidean'$, as it provided a better lobster-body result, also the $`manhattan'$ distance helps with reducing the impact of extreme outlier values, and provided faster processing of Density Based Clustering. The results of the selection Eq.~\ref{eq:model-lobster-body} are shown in Figure~\ref{fig:lobster-vetting}.  

\begin{figure} 
\includegraphics[width=\linewidth]{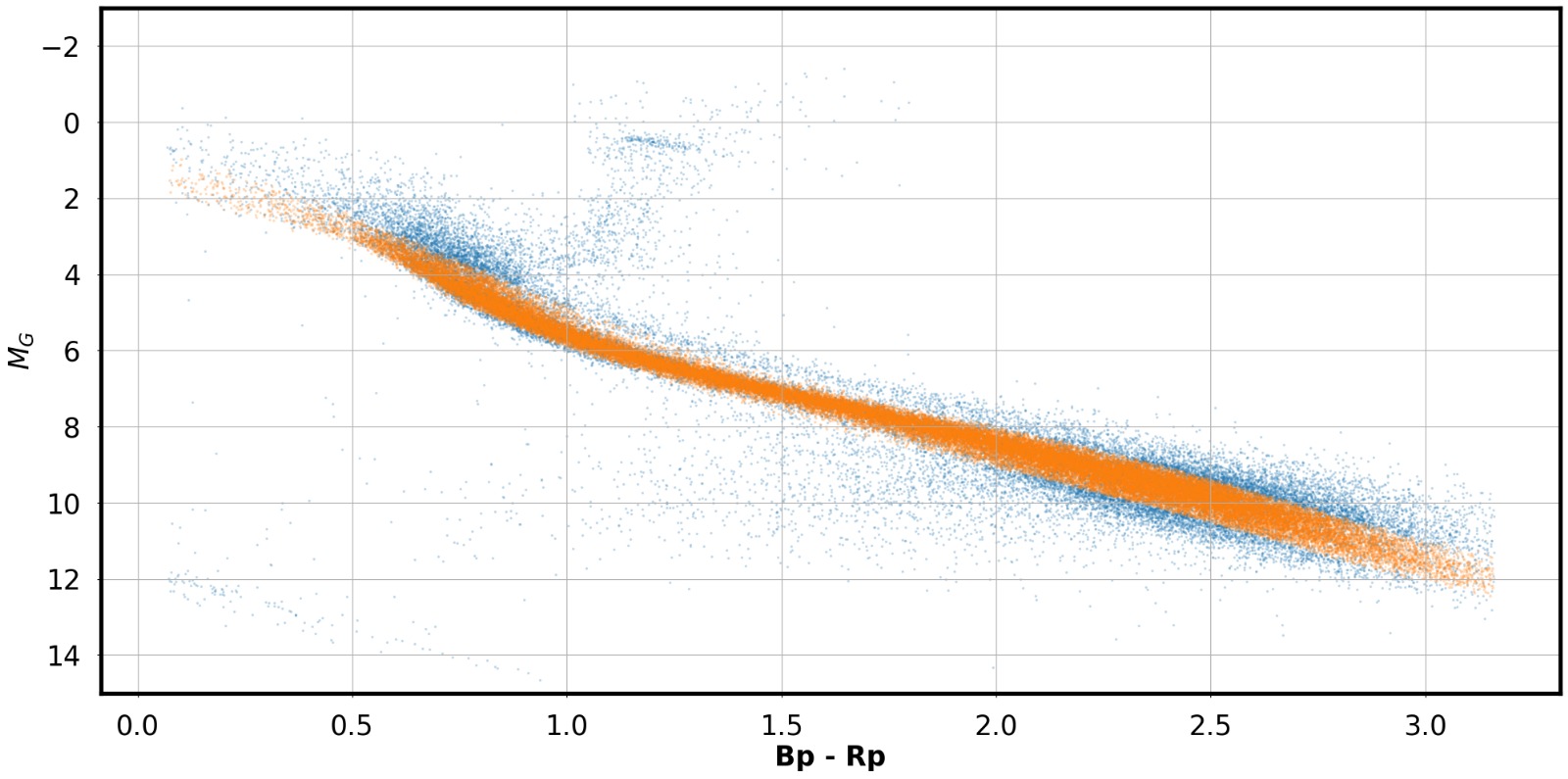} 
\caption{Colour-magnitude diagram of $M_G$ absolute magnitude vs $B_P - R_P$ colour; for both
 stars in each WB candidate. Blue points fail the lobster cut, orange points pass the lobster cut.
  \label{fig:cmd-lobster} 
} 
\end{figure}  

We then select the subset of WB-DR3-FLAME binary candidates with values of $( \dmgone, \dmgtwo )$ within the lobster-body defined above.  This reduces the sample to 23,223 wide-binary candidates; as this is the final part of our data cleaning, this sample is now our 'Cleaned' WB-DR3-FLAME (CWB-DR3-FLAME); this is the sample used exclusively in subsequent analysis.     The colour-magnitude diagram for both stars in these candidates is shown in Figure~\ref{fig:cmd-lobster}, showing that the lobster cut removes stars on the upper side of the main sequence, 
  and also removes a number of white dwarfs and subgiants, leaving a clean main sequence population.  
{\newa Our wider colour range and presence of white dwarfs and subgiants explains why our lobster diagram Figure~\ref{fig:lobster-vetting} has more outliers compared to \citet{Hartman_2022} which used more pre-selection cuts; but our
 lobster-body selection still gives a rather clean main sequence in Figure~\ref{fig:cmd-lobster} . }  

 \subsection{Results and scaled velocities} 
\label{sec:deltavplots} 	
\begin{figure*} 
		\begin{center} 
			\includegraphics[width=16cm]{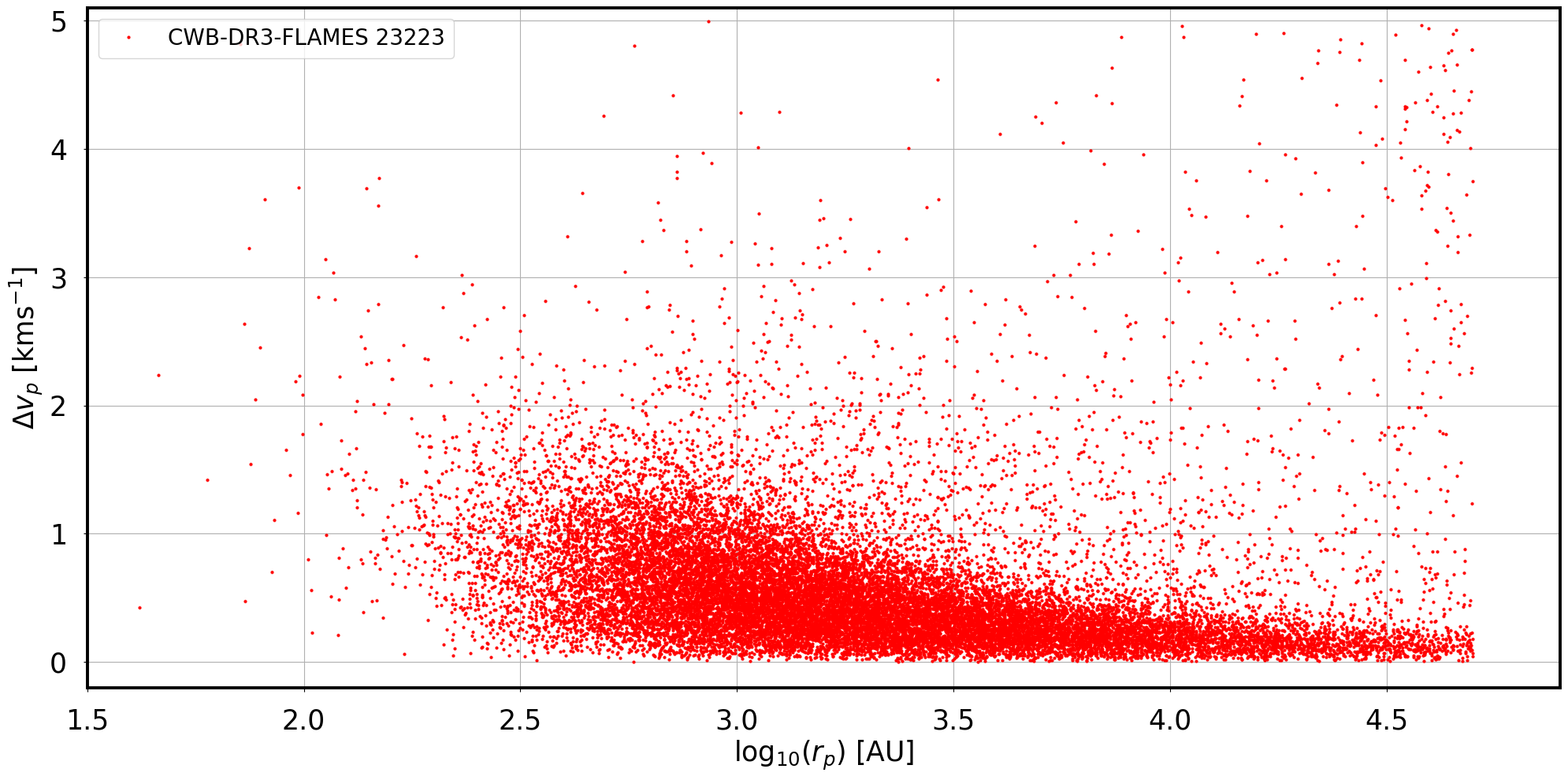} 
			\caption{Scatter plot of projected relative velocity $\dvp$ (y-axis)
				vs projected separation (log scale, x-axis) for the CWB-DR3-FLAME binary candidates.  
				The main selection cuts are visible at top and right. } 
			\label{fig:rpvp} 
			
                           \vspace{5mm}  % bit more whitespace between Figs. 
			\includegraphics[width=16cm]{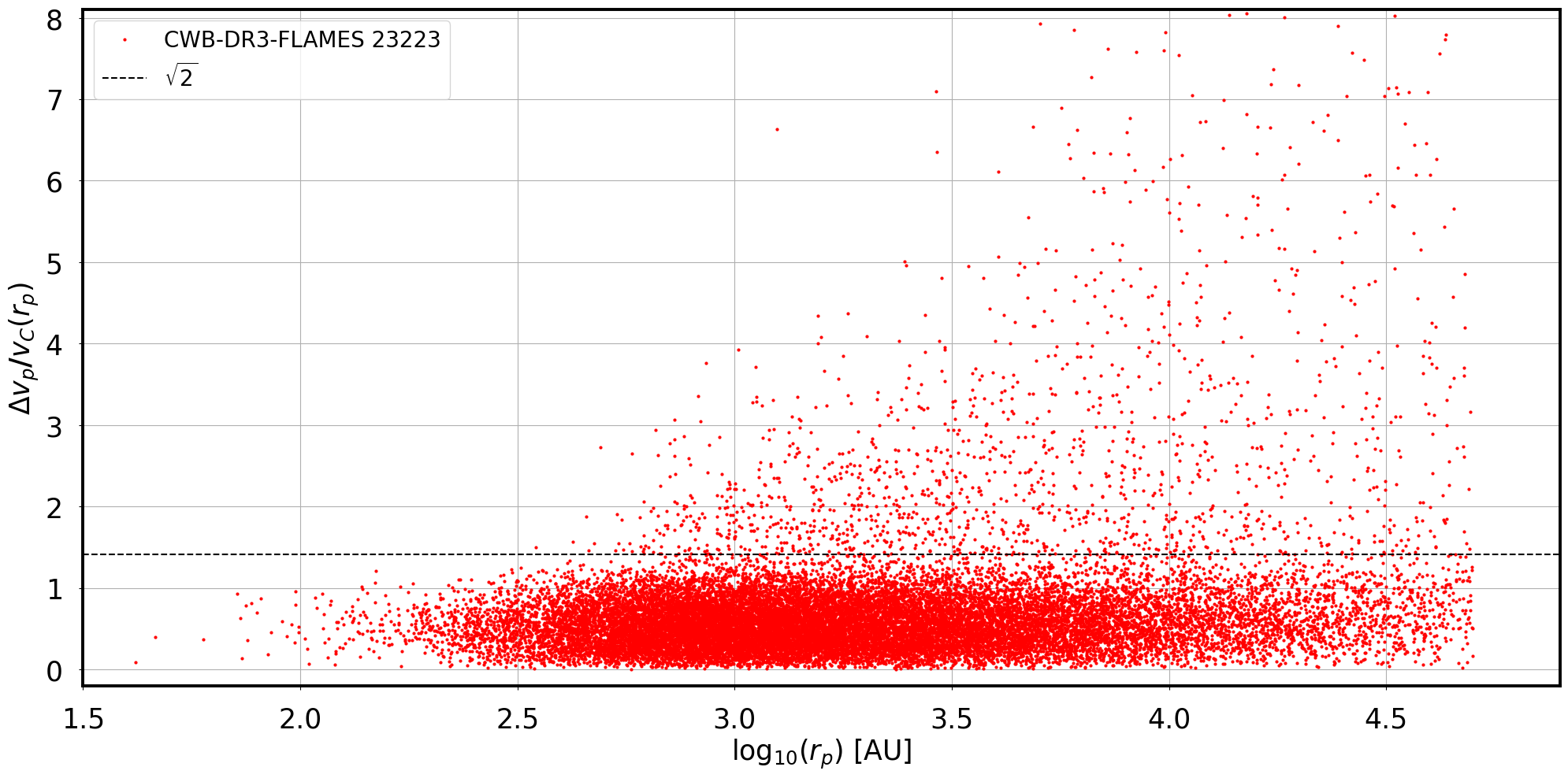} 
			\caption{Scatter plot of projected velocity relative to Newtonian, 
				$\vtilde$ from Eq.~\ref{eq:vtilde}, vs projected separation for CWB-DR3-FLAME sample.  
                                   The dashed line at 
			$\sqrt{2}$ indicates the Newtonian limit. The upper $\dvp$ cutoff now becomes 
                                   a diagonal curve, causing the empty region at upper left, but this border is 
				not sharp due to the additional dependence of $\vtilde$ on mass.  
				Note that in later analysis we only use the sample at $\rp > 1.25 \kau$ so $\log_{10} r_p > 3.09$, 
				so incompleteness in the upper-left region is unimportant.   }  
			\label{fig:rp_vratio}
		\end{center}
	\end{figure*}

	For the surviving 23,223 candidate binaries above, we show
	a plot of projected velocity difference vs projected separation 
	in Figure~\ref{fig:rpvp}; similar to \citetalias{Pittordis_2023}, this shows a clear excess at low 
  $\dvp$ approximately
	as expected for bound binaries, with an overdense cloud following
	a locus $\dvp \sim 1 \kms (\rp / 1 \kau)^{-0.5}$. 

%% New paragraph. 
     The population above the dense cloud is slightly sparser than the corresponding figure in \citetalias{Pittordis_2023}, indicating that the additional cuts are probably successful at removing
  many but not all higher-order multiples.   It is also notable that the points become sparser towards the top
 of the diagram at $\dvp > 2.5 \kms$,  except in the upper right corner at $\rp \ga 30 \kau$ so 
 $\log_{10} \rp \ga 4.48$ in Figure~\ref{fig:rpvp};  random pairs would
 produce the opposite trend, increasing with $\dvp$ due to phase-space volume; 
  this supports our conclusion below
 that random or unbound pairs are only a small contribution at $\rp < 20 \kau$, 
and the high-velocity tail is likely dominated  by triple/quadruple systems.   

	It is more informative to rescale to the typical Newtonian
	orbit velocity, so next we use the estimated masses described in ~\ref{sec:mass-estimate-flames} .
	%%% The key figs on velocity differences. 
	
	As in \citetalias{Pittordis_2023} for each candidate binary we then define 
	\begin{equation} 
		v_c(\rp) \equiv \sqrt{G M_{tot} / \rp} 
		\label{eq:vc} 
	\end{equation} 
	as the Newtonian circular-orbit velocity at the current 
	{\em projected} separation; 
	and we then divide the measured projected velocity 
	difference by the above to obtain a dimensionless ratio 
	\begin{equation} 
		\vtilde \equiv \frac{\dvp }{ v_c(\rp) } \ .  
                \label{eq:vtilde} 
	\end{equation} 
  This $\vtilde$ parameter is becoming commonly used in the recent literature, e.g. \citetalias{Pittordis_2023},   \citet{Banik_2024}, \citet{Chae_2024}.  By construction in standard gravity, $\vtilde = 1$ for a face-on circular orbit, 
 and $\vtilde < \sqrt{2}$ for a bound pure binary; also the median $\vtilde \sim 0.6$ for random 
  inclinations due to projection factors; see the following Section~\ref{sec:vtilde-stats} for details.  
	A scatter plot of $\vtilde$ vs $\rp$  for our cut sample is shown in 
	Figure~\ref{fig:rp_vratio}, and 
	 histograms of this ratio in bins of $\rp$  are compared with models below in Section~\ref{sec:fits}. 	
	
	\subsection{Transverse velocity errors} 
	\label{sec:vperrs} 
	We have estimated relative-velocity errors assuming
	uncorrelated errors between the two components of the binary, 
	simply from the root-sum-square of the quoted rms errors in $\mu_\alpha$
	and $\mu_\delta$ for each of the two stars in each binary, 
	and multiplying by mean distance to obtain the transverse-velocity 
	error. (This should be reasonable as long-range correlated errors should 
	mostly cancel between the two stars).  {\newa Parallax errors are ignored here
 as these contribute a multiplicative error $\la 1 \%$ to $\dvp$, much smaller than the 
  additive errors from proper-motion uncertainty}.

\begin{figure*}
		\begin{center} 
			\includegraphics[width=\linewidth]{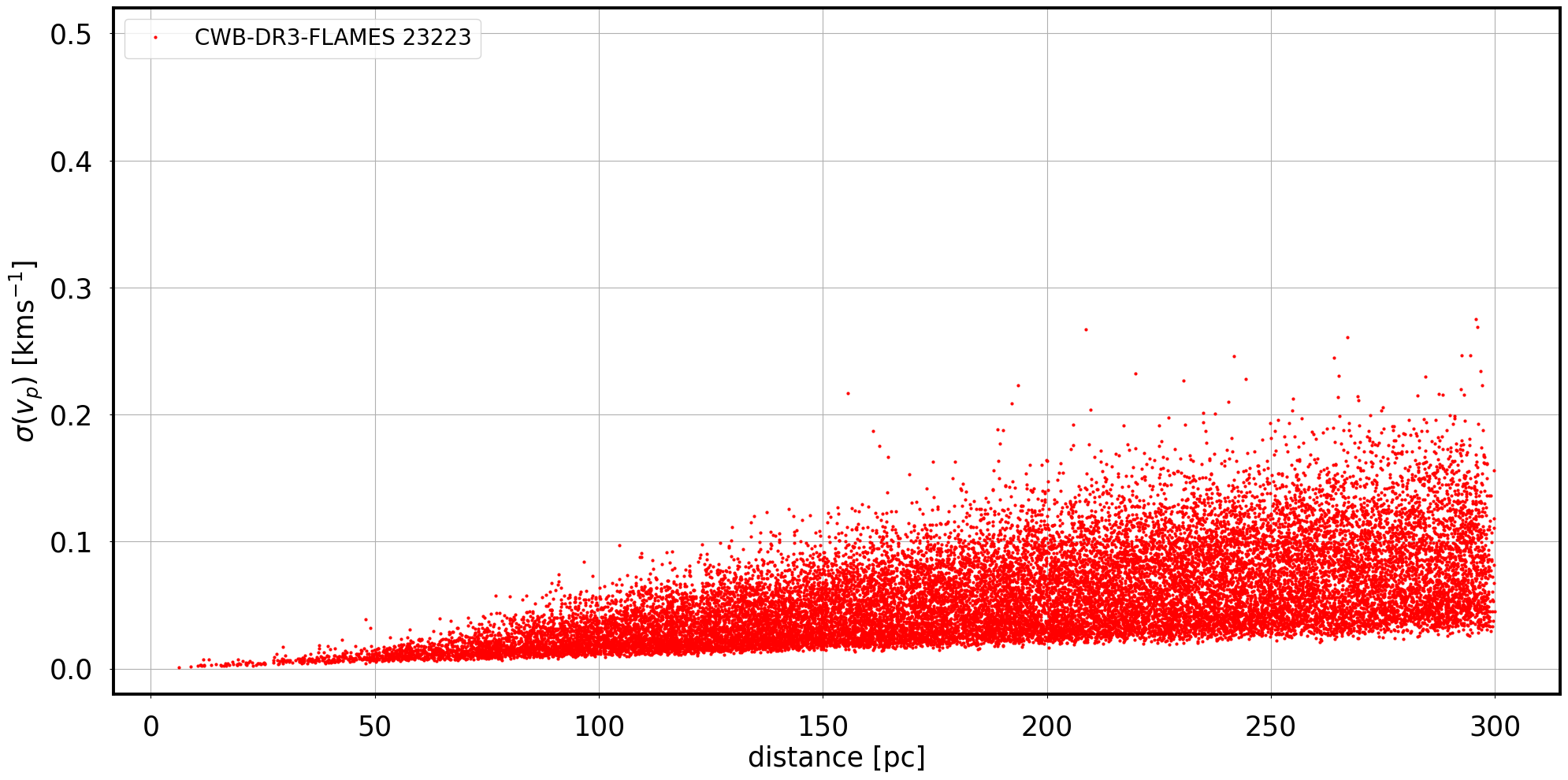} 
			\caption{Scatter plot of rms velocity uncertainty $\sigma(\dvp)$ versus
				mean distance for candidate binaries in CWB-DR3-FLAME.}
			\label{fig:dsigv} 
		\end{center}
	\end{figure*}

	Table \ref{tab:GDR2ToEDR3TransVelErrors} shows the comparison of the transverse velocity random errors between the CWB-EDR3 and CWB-DR3-FLAME. We can see using the FLAME masses and the extra `Lobster Body cut' has made {\newa a small improvement}, with the median at an impressively small value of $\sigma(\dvp) \approx 0.04 \kms$.
	When converting to the ratio to circular-orbit velocity $\sigma(\dvp)/v_c(\rp)$,  we see a negligible difference between the data sets. However, when reducing to the ``wide'' subsample with $5 < \rp < 20 \kau$, the
 values reduce by on average $ \sim 0.1$, with a median reduced to 0.08 and the 80th percentile is 0.16. 
	A scatter plot of $\sigma(\dvp)$ 
	versus distance is shown in Figure~\ref{fig:dsigv}; the trend
	with distance is clear, but
	most systems have $\sigma(\dvp) \la 0.1 \kms$ even near our
	$300 \pc$ distance limit.

	\begin{table}
		\caption{Table comparing percentiles of transverse velocity errors 
			and relative to circular velocity,  between CWB-EDR3 (PS23) and CWB-DR3-FLAME data sets. 
			Upper table for the full sets, lower table for subsample with $5 < \rp < 20 \kau$. }
		\label{tab:GDR2ToEDR3TransVelErrors}  
		\centering 
		\small
		\begin{tabular}{l c c} 
			%\hline
			\newline
			All $\rp$   &   CWB-EDR3   & CWB-DR3-FLAME    \\ 
			%\hline 
			\hline
			% data pasted from fit_vphist_alle_13may.log
			% Dont need median as already in next row. 
			%		\multicolumn{1}{|p{3cm}|}{\centering $\sigma(\dvp)$ \\ Median \\}
			%	& $\approx\,0.09$ &  $\approx\,0.06$   
			\\
			
			\multicolumn{1}{p{2.5cm}}{\centering $\sigma(\dvp) \  [\kms ]$ \\ (50\%, 80\%, 90\%) \\} & $\approx\,[0.06, 0.1, 0.13]$ &  $\approx\,[0.04, 0.08, 0.1]$   \\
			
			\multicolumn{1}{p{2.5cm}}{\centering $\sigma(\dvp)/v_c(\rp)$ \\ (50\%, 80\%, 90\%) \\} & $\approx\,[0.06, 0.12, 0.19]$ &  $\approx\,[0.06, 0.13, 0.2]$   \\
			\hline
			& & \\
			%\hline
			\multicolumn{1}{p{2.5cm}}{\centering ($5 < \rp < 20 \kau$)\\} 
			&  CWB-EDR3  & CWB-DR3-FLAME \\
			\hline
			\vspace{-0.1cm}
			\\
			\multicolumn{1}{p{2.5cm}}{\centering $\sigma(\dvp)\ [\kms ]$ \\ (50\%, 80\%, 90\%) \\} & $\approx\,[0.05, 0.09, 0.11]$ &  $\approx\,[0.05, 0.08, 0.1]$   \\
			
			\multicolumn{1}{p{2.5cm}}{\centering $\sigma(\dvp)/v_c(\rp)$ \\ (50\%, 80\%, 90\%) \\} & $\approx\,[0.14, 0.26, 0.34]$ &  $\approx\,[0.08, 0.16, 0.23]$   \\
			\hline
		\end{tabular}

	\end{table}

	% to go across both columns size of plot ##
	
	The latter values are significantly smaller than 1, but not very small, 
	so the effect of random proper motion errors will affect the detailed
	shape of the distributions below.  However, in future \GAIA
	data releases these values are expected to reduce by factors
	of at least 2--4 as  proper motion precision scales
	as $\propto t^{-3/2}$ for fixed scan cadence,  so the random errors in proper motions  
	are likely to become relatively unimportant in the medium-term 
	future.  
	
	We note that for a ``typical'' binary below 
	at $\rp \sim 5 \kau$ and $d \sim 200 \pc$,    
	the angular separation is 25 arcsec, so these are very well resolved and the uncertainty
	on $\rp$ is essentially the same as the error on the mean distance,
	typically below 1 percent and almost negligible.  
	The error on $\dvp$ is dominated 
	by random measurement errors on the proper motions, assuming that
	correlated systematic errors mostly cancel between the two components
	of each binary.   Since we are mostly interested in statistical
	distributions, the effect of random errors is modest as long
	as these are not larger than $\sim 0.2$ in $\vtilde$.  
	Note that for systems with small observed ratios 
	$\vtilde \sim 0.4$, the {\em fractional} uncertainty in this
	ratio is rather large; 
	however such systems still have a high probability
	of the true ratio being $\la 0.8$, so this scatter is relatively
	unimportant.  
 In the following fitting, it will turn out that the distribution between $0.8 \la \vtilde \la 1.5$ is the 
 most important discriminant between gravity models: 
     for these systems, the fractional
	uncertainty is relatively modest.   We include simulated proper-motion noise errors 
 in our model binary \& triple orbits as described below; this follows the recommendation of 
 \citet{Hernandez_CA_2024}. 

  While a possible tail of non-Gaussian errors in the \GAIA data remains a slight concern, this should improve in 
	future \GAIA releases as more observing epochs become available
	to reject outliers.   
   The precision will continue to increase with future DR4 and beyond in the extended mission, 
 so \GAIA random errors will become negligible
compared to other sources of uncertainty (especially contamination from triple systems, see below).

%% OOOOOO  orbits 
	\section{Orbit simulations for $\vtilde$ distributions} 
	\label{sec:orbits}  
  Our main  analysis below consists of fitting the joint distribution of candidate binaries in the $(r_p, \vtilde)$ plane 
 seen in Figure~\ref{fig:rp_vratio} to a mixture of simulated binary, triple and other populations; below 
 we will consider binaries in the range $1.25 \kau  \le r_p \le 20 \kau$.   Here the lower limit 
 ensures that angular separations of these WBs are $\ge 4$ arcsec, so WBs are well resolved and 
 minimising possible windowing effects   in the \GAIA data. 
 For the upper limit, beyond $20 \kau$ there are fewer systems, 
  perspective-rotation effects and proper-motion measurement errors become
   relatively larger, and binaries are more susceptible to disruption by encounters with other stars.  
  The selected range spans internal accelerations from $\sim  30\, a_0 $ to $0.2\, a_0$, so  any MOND  transition should be clearly observable, if present. 
% We divide this into 8 logarithmic bins 
% with each bin upper edge $\sqrt{2} \times $ the lower edge, so the first bin spans $1.25$ to $1.77 \kau$ and
% the eighth bin is $14.1$ to $20 \kau$; this gives the first two bins in the quasi-Newtonian regime where
% MOND effects should be small, bins 3 -- 5 are in a transition regime, while the final three bins are well into the 
%  MOND regime with internal accelerations below $a_0$.   We construct a histogram of $\vtilde$ as above, 
%  separately for each slice; 

 After binning, the observed histograms of $\vtilde$ in $\rp$ bins are fitted later 
 to an adjustable mixture of simulated binary, triple and other populations; 
 for the modelling, we construct simulated $\vtilde$ histograms for separate populations of pure-binaries, pure triples,  flybys and random projections; then in the fitting below, 
  the shapes of $\vtilde$ distributions for each population are held fixed at the simulation results, while the relative numbers of each population are adjusted to fit the data.
 In this section we summarise some properties of the $\vtilde$ distribution for Newtonian binaries
 in Section~\ref{sec:vtilde-stats}, then 
 describe the parameters used to generate the model populations in the simulations 
 in Sections~\ref{sec:orbsmg} -- \ref{sec:flybys}.   
 
	\subsection{Statistics of $\vtilde$ for pure binaries} 
	\label{sec:vtilde-stats} 	

	 A useful property of $\vtilde$ is that for pure binary systems and Newtonian gravity, 
	its statistical distribution can be robustly predicted given an assumed eccentricity distribution $f(e)$, 
	and the solid assumptions of random orbit orientations and random phases.  (We note that if orbit orientations are non-random, they will likely correlate with the Galactic plane; since our sample covers a wide range of Galactic latitude $15^o < \vert b \vert < 90^o$, this should mostly wash out any 
 Galactic alignment effects).  
	
	If we had access to 3D velocities and separations, we can define an analogous quantity 
	$\vtilde_{3D}  \equiv \vert \Delta \mathbf{v} \vert /v_c(r) $; as noted in \citetalias{Pittordis_2018} this is simply given by $\vtilde_{3D} = \sqrt{2 - (r/a)}$. The cumulative probability distribution for $\vtilde_{3D}$ is given by Eqs. 2 - 5
	of \citetalias{Pittordis_2018}, and analytic expressions for the 
	differential probability distribution are given by \citet{Benisty_2023}.   However, in practice this 3D version is not usefully observable: radial velocity  differences
	are measurable in principle, but very expensive in telescope time to reach the required 
 precision $\la 0.05 \kms$ for a large sample, while the radial component of binary separation is essentially impossible to measure to the required $\kau$ precision; so later we concentrate solely on the 2D $\vtilde$. 
	
	It was noted in \citetalias{Pittordis_2018} that the 90th percentile of $\vtilde_{3D}$ is close to 1.15 and is fairly
	robust against changing the eccentricity distribution $f(e)$;  here we show that in 2D there is an analogous result for $\vtilde$, 
	the numerical value is shifted downwards by projections since $\vtilde \le \vtilde_{3D}$, but 
	the 90th percentile of $\vtilde$ is robustly close to $0.94$ for a wide range of plausible eccentricity distributions. 
	
\begin{figure*}% [h]
\includegraphics[width=0.55\linewidth]{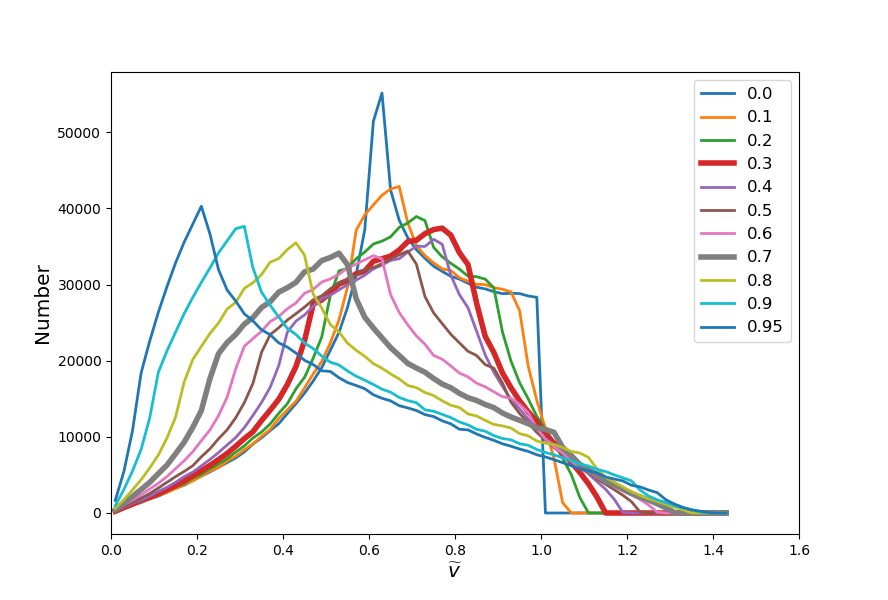} 
\hspace{-8mm}  % get rid of some whitespace in the middle, 
\includegraphics[width=0.55\linewidth]{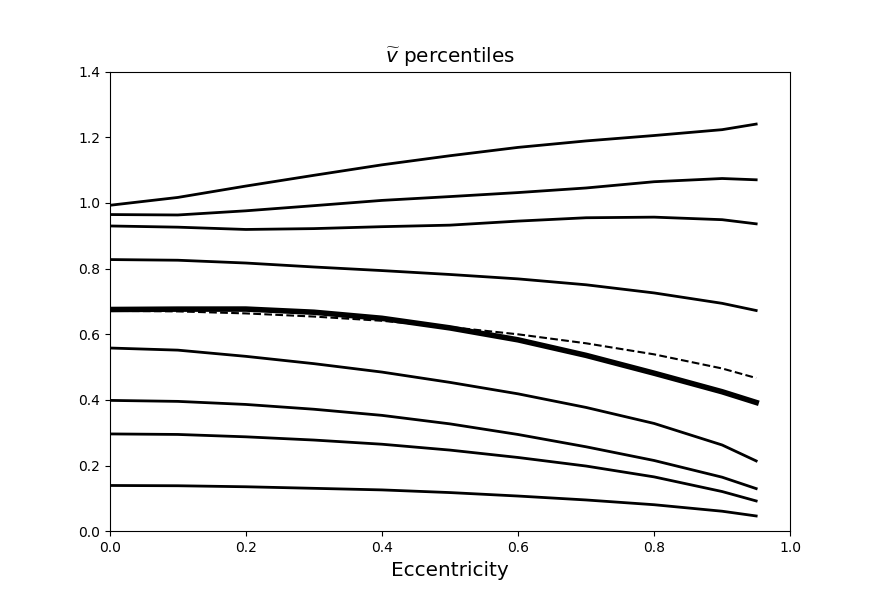} 
\caption{ Left panel: Probability distribution of $\vtilde$ for simulated binary orbits with fixed values of $e$, from 0.0 to 0.9 in steps of 0.1, also 0.95.   Values are as labelled in the legend; $e = 0.3$ and $0.7$ are thicker lines. \\
Right panel: Selected percentiles of $\vtilde$ distribution for orbits with fixed $e$ (abscissa). 
   Solid lines are percentiles 1 (bottom), 5,10, 25, 50 (thick line),, 75, 90, 95, 99 (top).  The dashed line is the mean.  
 \label{fig:vtilde_vs_e} 
}
\end{figure*}

	In Figure~\ref{fig:vtilde_vs_e} we show probability distributions for $\vtilde$ for fixed values of $e$, 
	and selected percentiles (1, 5, 10, 25, 50, 75, 90, 95, 99) of this distribution;  this would be the result
	for a single binary of fixed $e$ observed by many observers at random times and viewing angles. 
	In practice we observe many binaries at a single time and viewing angle, and 
	have no knowledge of $e$ for any single binary, so  
	the theoretical distribution is simply a sum of these weighted by a model for $f(e)$. 
	Note that for low-$e$ orbits there is a prominent caustic spike 
	near $0.62$; as in \citetalias{Pittordis_2019} this corresponds to the maximum of $\vtilde$ for nearly 
	edge-on low-$e$ orbits. As $e$ increases, the spike
	broadens into a ramp, and a tail appears extending to a maximum value
	of $\sqrt{1+e}$. The mode of the distribution initially increases with $e$, then decreases sharply for high $e$, so high-$e$ orbits produce many more systems at low $\vtilde \la 0.4$, fewer at intermediate values 
	$0.4 \la \vtilde \la 0.9$, but a similar number at $\vtilde > 0.9$.  This is due to the well-known property 
 that high-$e$ orbits spend most of the time moving slower than average at $r > a$, and a small fraction of time 
  moving faster than average at $r < a$. 
	For the cumulative percentiles, note that the mean and median of $\vtilde$ shift slightly
	downwards with increasing $e$; the upper 95th and 99th percentiles increase with $e$ as the tail 
	population at $\vtilde > 1$ increases, while as in 3D the
	90th percentile,  $\vtilde_{90}$, is relatively insensitive to $e$;  
	considering the variation of $\vtilde_{90}$ as a function of $e$, it shows a small  roughly sinusoidal variation  with a minimum 
	of $\vtilde_{90} = 0.920$ near $e = 0.25$, and a maximum of $\vtilde_{90} = 0.955$ near $e = 0.75$. 
	When averaging over a realistic smooth distribution $f(e)$, we get a robust prediction that 
	$\vtilde_{90} \simeq 0.94 \pm 0.01$ 	for pure binaries in Newtonian gravity, with minimal
  dependence on the uncertain $f(e)$.  

\begin{figure}[h] 
\includegraphics[width=1.1\linewidth]{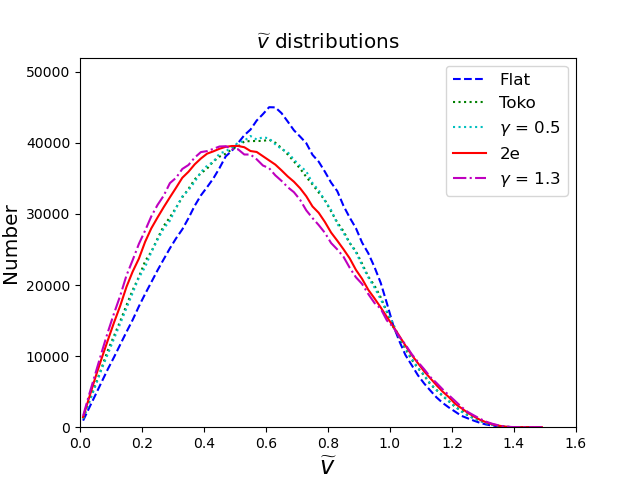} 
\caption{Probability distribution of $\vtilde$ for five eccentricity distributions $f(e)$: flat $f(e) = 1$ (dashed, blue); model $f(e) = 0.4 + 1.2e$ from \citet{Tokovinin_2016} (dotted, green); power-law model with $\gamma = 0.5$ (dotted, cyan); thermal distribution $f(e) = 2e$ (solid red); and power-law model with $\gamma = 1.3$ from \citet{Hwang_2022} (dot-dash, magenta).  Note that the $\gamma = 0.5$ and Tokovinin models give a very similar result.  
\label{fig:vtildemods} 
} 
\end{figure}

	In Figure~\ref{fig:vtildemods} we show the predicted distribution of $\vtilde$ for 5 different distributions $f(e)$: 
	the flat distribution $f(e) = 1$;  the distribution $f(e) = 0.4 + 1.2e$ fitted from \citet{Tokovinin_2016}; and three power-law models $f(e) = (1+\gamma) e^\gamma $ with $\gamma = 0.5, 1, 1.3$.  The $\gamma = 1$ model
	is equivalent to the ``thermal" distribution $f(e) = 2e$, while $\gamma = 1.3$ is close to the super-thermal 
	distribution favoured by \citet{Hwang_2022}.  
	In this Figure~\ref{fig:vtildemods}, we see that the flat $f(e)$ produces a pointed peak, resulting from the caustic spike from low-$e$ orbits.  As the distribution is weighted more to higher $e$, the $\vtilde$ histograms
	become more rounded, the mean and mode shift to lower values, but the skewness increases and the upper tail grows slightly.  The Tokovinin and $\gamma = 0.5$ models, though different, produce very similar $\vtilde$ distributions.  There is a pivot point where all models nearly intersect near $\vtilde \simeq 1$, so the total fraction 
above $\vtilde \ge 0.9$ is nearly constant:  
  this fraction is between 12.6 and 12.8~percent for all five $f(e)$ distributions above.  

	Later it is also useful to consider the fraction of pure binaries giving $\vtilde \ge 0.8$: 
  this is slightly more sensitive to the $f(e)$ distribution, but offers a larger percentage for improved statistics, and a larger ratio of binaries to triples in the modelling,  
	hence is less sensitive to details of the triple modelling. 
  For fixed $e$ this fraction at $\vtilde \ge 0.8$ monotonically declines from 29.2 percent at $e = 0$ to 16.8 percent at $e = 0.95$; however, when averaging over a reasonable $f(e)$ distribution this variation 
 is substantially reduced, 
	giving 23.2 percent for flat $f(e)$, 21.9 percent for both Tokovinin and $\gamma = 0.5$ models, 
	21.0 percent for $f(e) = 2e$, and 20.6 percent for the $\gamma = 1.3$ model.   
	
	In our example MOND model (see below), though numerical orbits are not closed, each single orbit
	at separation $\ga 10 \kau$ is roughly  
	approximated by a Keplerian orbit with an effective $G$ boosted by a factor $1.36$, and velocities boosted
	by a factor 1.17 ; our definition of $\vtilde$ has the Newtonian $v_c$ on the denominator, 
	thus the predicted $\vtilde$ distribution in MOND is very similar to the Newtonian prediction rescaled 
	by this factor of 1.17 multiplicative shift. 

	As a result of this, the MOND models produce a dramatic increase in 
  fractions of binaries above $\vtilde$ thresholds of 
	$0.8$ or 0.9:  the MOND model with $f(e) = 2e$, for binaries wider than $r_p \ge 10 \kau$,  
	predicts 33.7 percent at $\vtilde \ge 0.8$  and $24.4$ percent above $\vtilde \ge 0.9$.  These are substantially 
   larger fractions compared to Newtonian predictions above, which {\bf cannot be reproduced in a Newtonian model}  with {\bf any}   eccentricity distribution. 
	Therefore, if we had a sample of $\ga 1000$ ``pure" wide binaries, we could readily distinguish between
	Newtonian gravity and MOND by simply testing whether $\vtilde_{90}$ is near 0.94 or 1.10,
  or testing whether the fraction above $\vtilde > 0.9$ is near 12 percent vs 24 percent; these would
 be easily distinguishable at high significance.    
	
	However, we do not currently have a sample of ``pure" binaries: the observed distribution of $\vtilde$ shows
	an extended tail up to much larger values $\vtilde \ga 4$, as originally noted by \citetalias{Pittordis_2019} in \GAIA DR2 data, and confirmed in \citetalias{Pittordis_2023}.  
	Although MOND does allow bound binaries with $\vtilde$ values above the Newtonian limit of $\sqrt{2}$, 
	values above $\vtilde \ge 1.75$ cannot be achieved in a reasonable MOND model with external field effect, and values	above 1.5 are very rare in our MOND simulations (frequency $\sim 0.1\,$percent);
  thus this extended tail is almost certainly not pure binaries, 
  but is dominated by triple, higher-order or unbound systems. 
	In Section~\ref{sec:fits} later, we fit the observed $\vtilde$ distributions as a mixture of simulated binary, triple and unbound systems in both Newtonian and MOND gravity, 
	with the $\vtilde$ distributions for model systems generated as described in the following subsections, 
 then relative abundances allowed to vary in the fitting in Section~\ref{sec:fits}.    
	
	\subsection{Binary orbit simulations} 
	\label{sec:orbsmg} 
	Similar to Section 3 from \citetalias{Pittordis_2023}, 
	here we simulate a large sample of $\sim 5 \times 10^6$ orbits
	in both Newtonian gravity and 
	a specific modified-gravity model; as before we use the model from \citet{Banik_2018_Centauri} (hereafter \citetalias{Banik_2018_Centauri}). 
		
	For the MOND models, the orbits are 
	not closed and not strictly defined by 
	the standard Keplerian parameters $a, \, e$. As in \citetalias{Pittordis_2023}, 
    we parametrise using the ``effective'' orbit size $\hat{a}$ and quasi-eccentricity $\hat{e}$ as follows: 
	we define $\hat{a}$ to be the separation at which the simulated relative velocity
	is equal to the circular-orbit velocity (in the current 
	modified-gravity model), then we define $\theta_{\rm circ}$ to be the angle between the relative velocity vector and the tangential direction when the orbital separation crosses
	$\hat{a}$, and then $\hat{e} \equiv \sin \theta_{\rm circ}$;  these definitions 
	coincide with the usual Keplerian $a, e$ in the case of standard gravity.  
 
   In both gravity cases, we simulate orbits with a 
 % flat distribution of $\log_{10} \hat{a}$ 
% updated with a^-1.6 and refs 
 {\newa distribution function
\footnote{This turns out to be the most significant 
 change from v1 of this paper; see discussion later in Sec~\ref{sec:caveats}.  }
  $dn/d\hat{a} \propto \hat{a}^{-1.6}$ for $\hat{a}$  between }  
  $0.3 \kau \le \hat{a} \le 150 \kau$ (much wider than the data sample to avoid edge effects); this -1.6 
 power law is consistent with observations, e.g. \citet{Lepine_2007}, \citet{Andrews_2017}, \citet{Badry_2018}. 
   We use a $\gamma = 1.3$ distribution for $\hat{e}$.

	The only modified-gravity model we consider below is the MOND model with ExFE, using the approximation 
	of \citetalias{Banik_2018_Centauri}; this is given by 
	\begin{eqnarray} 
		g_{N,int} & = & G(M_1 + M_2)/r^2  \\
		g_{N,gal} & = & 1.2 \, a_0 \\ 
		g_{N,tot} & = & \left( g_{N,int}^2 + g_{N,gal}^2 \right)^{1/2} \\ 
		g_{i, EFE} & = & g_{N,int} \nu(g_{N,tot}/a_0) \left(1 + 
		\frac{\kappa(g_{N,tot})}{3} \right) \\  
		\kappa & \equiv & \frac{\partial \ln \nu }{\partial \ln g_{N} } 
		\label{eq:exfe} 
	\end{eqnarray}
	where $g_{N,int}$ is the internal Newtonian acceleration of the binary; 
	$g_{N,gal}$ is the external (Galactic) Newtonian acceleration, 
	$g_{N,tot}$ is the quadrature sum of these, $\nu$ is the MOND $\nu$ function 
         defined in \citet{McGaugh_Lelli_2016} which produces a good fit to spiral galaxy rotation curves,   
  which is 
\begin{equation} 
\label{eq:mls} 
  \nu(y) = \frac{1}{1 - \exp(-\sqrt{y}) }  \ , 
\end{equation}  
	and $g_{i, EFE}$ is our model MOND-ian 
	internal acceleration, approximating the 
	application of the external field effect. 
	(This is not quite an exact solution of the MOND-like equations,
	but is shown by \citetalias{Banik_2018_Centauri} to be a good approximation to the full 
	numerical solution).  
%  It also provides a good fit to Galactic rotation curves as noted in  \citetalias{Pittordis_2023}. 
	
%	Above,  the observed  Galactic rotation values 
%	$v_{LSR} \simeq 232 \kms$ and $R_0 \simeq 8.1 \kpc$ imply 
%	a total Galactic acceleration close to $1.75 \, a_0$, hence we 
%	require $g_{N,gal} \nu(g_{N,gal}/a_0) \approx 1.75 \, a_0$. Solving this 
%	leads to $g_{N,gal} \approx 1.16 \, a_0$ as above and 
%	$\nu \approx 1.51$, in reasonably  
%	good agreement with the estimated baryonic contribution to the Galactic
%	rotation (as expected, since the MLS fitting function was derived by 
%	fitting to a sample of external spiral galaxies with well-observed rotation
%	curves, so this is consistent with our Galaxy being typical).  

\subsubsection{Model perspective effects} 
\label{sec:perspec} 
 We also model perspective effects on the velocity differences, as in Eq.~6 of \citetalias{Pittordis_2018}, multiplied  below by $d$ to convert proper motion to velocity units.  
Perspective effects are modelled as follows: 
  for each simulated observation,  we model the barycenter velocity as an isotropic Gaussian with a 
 velocity dispersion of $25 \kms$ in each coordinate.   Given the simulated distance and random viewer direction, 
  we calculate the barycenter radial velocity $v_r$, transverse velocity $v_{tan}$, and radial separation $\Delta d$ 
  and angular separation $\theta$.  
 We then add a term $\Delta v_{shrink} = - v_r \, \theta$  (for $\theta$ in radians) parallel to the angular separation  direction for the apparent contraction of
 the binary due to its radial velocity, and another term $\Delta v_{ov} = v_{tan} (\Delta d / d) $ for the closer 
 component apparently overtaking the more distant one in the direction parallel to $\mathbf{v}_{tan}$; these are added as vectors to the simulated orbital velocity difference $\Delta\mathbf{v}_P$ before converting to $\vtilde$.  
  These perspective terms turn out to make only a minor change to the overall results; the effect can 
 become  significant for wide separations $\rp \ga 10 \kau$ and small distances $\la 100 \pc$, but 
 these systems are a small fraction  of our total sample.

\subsubsection{Model proper motion errors} 
\label{sec:vtilde-noise} 	
In the binary-orbit simulations, we add simulated measurement noise to the theoretical $\vtilde$ 
 values as follows: we use simulated $G$ magnitudes for both stars input to 
 the fitting formula of \citet{Kluter_2020} for the rms position error per single \GAIA scan,  
\begin{eqnarray} 
	\label{eq:sigast} 
	\sigma_{AL} & = & \frac{ 100 + 7.75\,  u  }{ \sqrt{9} } \, \mathrm{microarcsec}  
          \\
      u & \equiv & \sqrt{ -1.631 + 680.766 \, z+ 32.732 \, z^2} \nonumber \\ 
	z & \equiv & 10^{0.4 \,[max(G, 14) - 15] }  \nonumber  
\end{eqnarray} 
%%  add ref.. arxiv:1911.02584 . 
where $\sigma_{AL}$ is the 1D along-scan astrometric precision for a single focal plane transit, crossing 9 CCDs. 

  Evaluating Eq.~\ref{eq:sigast} vs magnitude, then comparing with the median {\newa proper motion errors} vs magnitude from Table~4 of \citet{Lindegren_2021}, we 
 find {\newa by inspection that rescaling $\sigma_{AL}$ by a multiplier of $0.26 /$year } produces a good approximation to the median observed proper motion errors in that Table:  so we adopt model proper motion 
 errors of $0.26 \, \sigma_{AL} \, \mathrm{yr}^{-1} $  per coordinate for each star, {\newa using its simulated magnitude in Eq.~\ref{eq:sigast}.  }      
 These values for both simulated stars are added in quadrature, and multiplied by distance then 
 divided by circular-orbit velocity to give an equivalent scatter in $\vtilde$. 
   We generate independent Gaussians for the error components parallel and perpendicular to the error-free 
 $\vtilde$,  and take the magnitude of the resultant 2D vector as the simulated noisy $\vtilde$.  
\vspace{5mm}   % avoid orphan title 

	\subsection{Triple system simulations}
	\label{sec:triple-sim} 	
	To simulate a population of triple systems, we generate two binary orbits made up of three stars. We choose labels so star 1 is the single star in the wide system, and stars 2 \& 3 comprise the inner binary, so the outer orbit is star 1 orbiting the barycentre of stars 2 and 3. 
	For the masses, for stars 1 and 2 we pick masses from a \citet{Kroupa_2013} IMF distribution,
 which is then multiplied by $(M/0.95 \msun)^{-2.5}$ for $M > 0.95 \msun$ to give a present-day main-sequence mass function  assuming constant past star-formation rate. 
 We then pick a random distance with distribution $\propto d^2$ within 300 pc; we then compute simulated
 apparent $G$ magnitudes, if both are $G < 17$ we accept the pair, otherwise discard and re-select. 
 Star 3 is then assigned $M_3 = q \,M_2$ with $q$ uniform in the range $[0.02, 1]$. 	

For orbit sizes, we choose the outer orbit size $\hat{a}_{out}$ with a {\newa power-law 
 distribution } $p(\hat{a}_{out}) \propto \hat{a}_{out}^{-1.6}$ within the range  $0.3 \kau \le \hat{a}_{out} \le 150 \kau$), as for the pure binaries above;  and the inner orbit size $a_{inn}$ is chosen from the lognormal distribution
  given by  \citet{Offner_2023} for FGK stars; this is a lognormal where $\log_{10} a_{inn}$ is a Gaussian with mean 
 $\log_{10} (40 \au)$ and standard deviation 1.5 ; we also apply an upper limit to $a_{inn} $ for stability, 
 based on the fitting formula of \citet{Tokovinin_2014} as 
 \begin{equation} 
 \frac{a_{inn} }{ \hat{a}_{out}} \le \mathrm{max}( 0.342 (1 - \hat{e}_{out})^2 \, , 0.01) 
\label{eq:ainnmax} 
\end{equation} 
 
For eccentricities, our default model uses the power-law distribution of \citet{Hwang_2022} 
 which is $f(e) = (1/2.3) e^{1.3}$ 	for the outer orbit;  and the 
 linear model from \citet{Tokovinin_2016}, which is $f(e) = 0.4 + 1.2e$ for the inner orbit; 
 inner and outer eccentricities are uncorrelated from these distributions.  
	
	For simplicity, we treat the inner and outer orbits as independent.   
	We solve for the two orbits independently in their own planes, and then apply a random 3D rotation matrix,
	$\mathbf{R}$,  to the inner orbit relative velocity to generate a random relative alignment between the two orbits. 
	Next the system is ``observed" at $\sim 10$  random phases and 5 random viewing directions 
 for each phase. At each simulated
	``observation" we evaluate the projected separation, and the 3D velocity difference between star 1 and 
	the ``observable center" of stars 2+3, as follows. 
\begin{equation} 
\label{eq:v3dobs}
	\mathbf{v}_{3D,obs} = \mathbf{v}_{out} - f_{pb} \, \mathbf{R}\mathbf{v}_{inn}
\end{equation}
where $\mathbf{v}_{out}$ is the outer orbit velocity (star 1 relative to the barycentre of 2+3), and $\mathbf{v}_{inn}$ is the relative velocity between stars 2+3.   
%% new bit year for 3-year average velocity. 
  In a refinement of \citetalias{Pittordis_2023}, instead of calculating the instantaneous velocity difference for
 $\mathbf{v}_{inn}$, we take the vector position difference at two epochs separated by 34 months, and divide
  this vector by 34 months, to obtain the time-averaged velocity over the baseline of the \GAIA DR3 data; this accounts approximately for orbit-wrapping for short-period orbits.  

 This velocity $\mathbf{v}_{inn}$ is rotated and scaled down by a dimensionless factor
	$f_{pb} \le 0.5$, defined as the fractional offset between the ``observable centre" and the barycentre of stars 2+3, 
	relative to their actual separation.  
	The ``observable centre" is defined according to the angular separation: if less than 1 arcsec, we assume stars 2+3 are
	detected by \GAIA as a single unresolved object, and use the luminosity-weighted centroid (photocentre) of the two. 
	Otherwise for separation $> 1\,$arcsec, we assume stars 2 and 3 are detected as separate objects, or star 3 is unobservably faint, and we take the position of star 2 alone as the observable centre.  Therefore, this results in a fractional offset given by 
	\begin{equation} 
		f_{pb} = \begin{cases} 
			\frac{M_3}{M_2+M_3} - \frac{L_3}{L_2 + L_3}  \ (\theta < 1 \text{ arcsec}  ) \\ 
			\frac{M_3}{M_2+M_3}   \ (\theta \ge 1 \text{ arcsec} ) 
		\end{cases} 
	\end{equation}  
	where the $L_{2,3}$ are the model luminosities. 
  
 The 3D resultant velocity Eq.~\ref{eq:v3dobs} above is then converted to 2D projected 
 velocity according to each 
 random viewing direction, and observables including $r_p, \vtilde$ are saved to create simulated histograms for triple systems; this procedure is repeated for both of the gravity models (Newtonian and MOND). 

We also include the "apparent mass" bias for triples: the actual outer-orbit velocity depends on the 
 sum of all three masses, while in calculating $\vtilde$ the mass of star 2 is estimated from the combined
 luminosity $L_2 + L_3$; this $M_{2,est}$ is smaller than the true $M_2 + M_3$, so the calculated $\vtilde$ is boosted
 by a multiplicative factor of $\sqrt{ [(M_1 + M_2 + M_3)/(M_1 + M_{2,est}) ]} $ , compared to the value 
 using the unknown true mass sum.  

\subsubsection{Simulating the triple cuts} 
\label{sec:sim-triplecuts} 
For a realistic triple simulation, we also need to include the effect of the cuts on {\tt ruwe}, faint companions, 
 {\tt ipd\_frac\_multi\_peak} and the Lobster diagram applied to our data sample above. These cuts selectively remove certain regions of triple parameter space, depending on the separation and 
  mass ratio of the inner binary,  so surviving triples have a somewhat different distribution of $\vtilde$ to 
 the original population.     The {\tt ruwe} cut preferentially removes 
  triples with inner-orbit period in a broad window around 3 years, while it is insensitive to 
 very small or very large orbits. The 
 {\tt ipd\_frac\_multi\_peak} cut preferentially removes inner angular separations $\ga 0.3$ arcsec, 
 while the Lobster cut removes inner pairs with roughly similar luminosities, independent of separation.  
 
For the {\tt ruwe} cut, we simulate a {\tt ruwe} value for the inner orbit: 
 the simulated $G$ magnitude of star 2 is used as input to the fitting formula 
  of \citet{Kluter_2020}, Eq.~\ref{eq:sigast} above. 
 We then approximate the rms dispersion of the photocentre of stars 2 + 3 from the barycentre over a full
 orbit as  
 \begin{equation} 
 \sigma_{cen} \equiv f_{pb} \, a_{inn} / (2d)
\label{eq:sigcen} 
\end{equation} 
  where the factor of 2 accounts for 
 $1/ \sqrt{2}$ for \GAIA's 1D scanning and a second $1/\sqrt{2}$ approximates projection factors; 
  we then adapt the fitting formula  from \citet{Penoyre_2022} and define  the simulated {\tt ruwe} value as 
\begin{eqnarray} 
\label{eq:ruwe-sim} 
 {\tt ruwe}_{sim} & \equiv & \frac{ \sqrt{ n_{orb}^4 \sigma_{cen}^2 + \sigma_{AL}^2 } } { \sigma_{AL} } \\ 
  n_{orb} & \equiv & \mathrm{min}(1, T/ P_{inn} )  
\end{eqnarray} 
 where the $n_{orb}$ factor accounts for the suppression of the residuals for long-period orbits with $P_{inn} \ge T$ 
  where only a partial arc is observed, and the linear term is absorbed into the \GAIA proper motion fit.  
 We then reject simulated systems with ${\tt ruwe}_{sim} > 1.2$ .  

For the {\tt ipd\_frac\_multi\_peak} cut, we simulate this by rejecting triples where the inner binary 
 has angular separation $> 0.3\,\mathrm{arcsec}$ and magnitude difference $< 4 \,\mathrm{mag}$. 

For the Lobster cut, we reject simulated triples when the inner binary is  more than 0.4 mag above the main sequence ridge-line, which corresponds to a mass ratio $M_3 / M_2 > 0.8$.  

%% para updated with values from 31 May runs 

\begin{figure*} 	
\includegraphics[width=\linewidth]{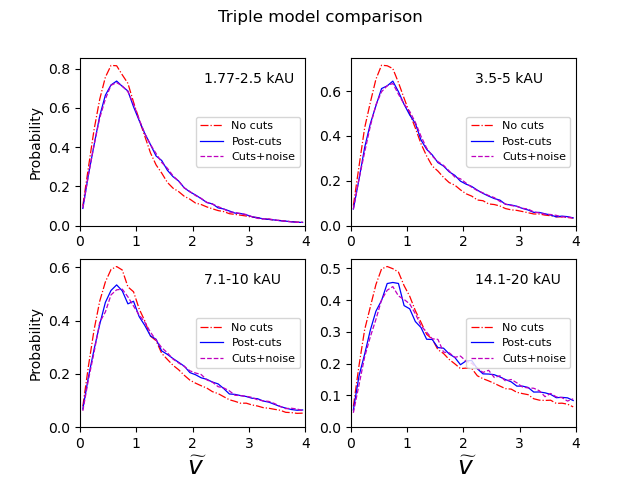}  %% updated v2 with 20250531 sims. 
\caption{Histograms of $\vtilde$ for our baseline triple simulation, in 4 selected bins of outer orbit $\rp$ 
 as labelled in the legend.  (Every second $\rp$ bin is shown for brevity).  Red dot-dashed line: before cuts. Blue/solid line: after cuts, noise-free,  renormalised to unit area. Magenta dashed line: after cuts with simulated $\vtilde$ noise. 
\label{fig:vtilde-trips} 
} 
\end{figure*}

The combined cuts above reject 61.3 percent of our simulated triple systems, leaving 38.7 percent
 surviving all the cuts.  The {\tt ruwe} cut is the most effective, rejecting 42\% of the original sample; the Lobster cut rejects a further 11 \%; while the {\tt ipd\_frac\_multi\_peak} cut and faint-companion cuts together reject 8 \%. 
 Figure~\ref{fig:vtilde-trips} shows the distribution of $\vtilde$ for triples pre- and post-cuts, and with 
 proper motion errors added as in Section~\ref{sec:vtilde-noise} above.  The survival
 fraction is a gently rising function of $\vtilde$, from $\approx 34\%$ at small $\vtilde$ to $\approx 46\% $ 
 at large $\vtilde$.  The survival fraction is weakly dependent on $\rp$, as expected 
 since the cuts depend on inner-orbit parameters, not outer-orbit values.  
     The effect of proper motion errors is nearly negligible
 here, as the intrinsic width of the triple distribution is much broader than the errors. 

Some selected summary statistics of the $\vtilde$ distribution for triples are listed in Table~\ref{tab:tripstats}; these
 may be useful for comparison with future work with data or other triple models. 
 These simulated triple distributions in the $\rp, \vtilde$ plane are then used (with fixed shape and variable normalisation) in the fitting procedure below in Section~\ref{sec:fits}.      

\begin{table*} 
\caption{ Summary statistics of $\vtilde$ for the baseline triple model, after simulated cuts, in bins of projected separation $\rp$.  Columns 2--4 are percentages 
 of triple systems (in this $\rp$ bin) in selected $\vtilde$ ranges; columns 5--6 are median and 90th percentile of $\vtilde$. \label{tab:tripstats}  }   
\begin{center} 
\begin{tabular}{l|ccc|cc} 
%%% Table values now updated with _31mar  triple model, after data cuts, no added noise. 
& \multicolumn{3}{c}{Percentage in $\vtilde$ range:} & \multicolumn{2}{|c}{Percentiles of $\vtilde$: } \\ 
$\rp$ bin (kAU) &   $\vtilde < 0.8$ & $\vtilde \in (0.8, 1.5)$ &  $\vtilde > 1.5$ &  Median  & 90th percentile \\
\hline  
%% Following 8 lines are old v1, 04-apr-25 triples... 
% 1.25 -- 1.77   &  44.6  & 35.8 & 19.6 & 0.87 & 1.96 \\ 
% 1.77 -- 2.5   &  41.1  & 34.6 & 24.3 & 0.94 & 2.22 \\ 
% 2.5 -- 3.5     &  37.6  & 33.4 & 29.0 & 1.00 & 2.51 \\ 
% 3.5 - 5.0    &  33.7  & 31.9 & 34.3 & 1.10 & 2.90 \\ 
% 5.0 - 7.1     &  31.1  & 29.1 & 39.8 & 1.20 & 3.37 \\ 
% 7.1 - 10.0    &  28.0  & 27.7 & 44.3 & 1.32 & 3.86 \\ 
% 10.0 - 14.1   &  25.7  & 25.6 & 48.6 & 1.45 & 4.43 \\ 
% 14.1 - 20.0   &  23.6  & 23.4 & 52.9 & 1.62 & 5.16 \\  

%  Table now updated with 31-May values . 
1.25 -- 1.77   &  44.6  & 35.3 & 20.0 & 0.87 & 1.98 \\ 
1.77 -- 2.5   &  41.5  & 33.9 & 24.6 & 0.93 & 2.24 \\ 
2.5 -- 3.5     &  37.9  & 33.1 & 29.0 & 1.00 & 2.51 \\ 
3.5 - 5.0    &  34.9  & 31.2 & 34.0 & 1.08 & 2.88 \\ 
5.0 - 7.1     &  32.5  & 28.7 & 38.8 & 1.17 & 3.34 \\ 
7.1 - 10.0    &  29.2  & 26.5 & 44.2 & 1.31 & 3.85 \\ 
10.0 - 14.1   &  26.2  & 24.9 & 48.9 & 1.46 & 4.42 \\ 
14.1 - 20.0   &  23.1  & 23.0 & 53.9 & 1.66 & 5.16 \\  
\end{tabular} 
\end{center} 
\end{table*} 

Histograms of $\vtilde$ for triples in some selected outer-orbit $\rp$ bins are shown in Figure~\ref{fig:vtilde-trips}, and selected summary statistics for each $\rp$ bin are listed in Table~\ref{tab:tripstats}. 
  The triple histograms show a peak around $\vtilde \sim 0.7 - 0.9$, increasing slightly with $\rp$; this peak is only moderately  shifted from the location $\vtilde \approx 0.55$ for pure binaries; 
 the relatively modest shift is due to common cases where the inner binary produces
 only a small velocity perturbation, due to orbit-averaging for short periods, and/or small $f_{pb}$ 
 for mass ratios near 0 or 1. Unlike binaries, the triple distributions show a long tail 
 extending well beyond $\vtilde \sim 2$; 
 the tail becomes more extended at larger $\rp$,   due to the effective $\sqrt{\rp}$ factor in $\vtilde$. 

The length of the tail can be approximately explained because 
  two effects above produce an effective ceiling on the magnitude of $f_{pb} \, {\bf v}_{inn}$: 
  considering for simplicity fixed masses, a circular orbit and varying period $P$, 
 the change of separation vector $\mathbf{r}_{23}$ over a time baseline 
  $T$ is $\Delta\mathbf{r}_{23} \propto P^{2/3} \sin(\pi T/P)$.  
 The global maximum of this occurs for $P = \pi T / u$ where $u = 0.9674$ is the smallest positive solution of $\tan u = 3u /2$, giving $P \simeq 3.25 \, T$.   Taking an example inner binary of  $0.8 + 0.4 \msun$ and $T = 34$ months for DR3, 
 this gives a  maximum time-average velocity $\Delta\mathbf{r}_{23} / T $  of $12.8 \kms$.  
 Also, the maximum of $f_{pb}$ for an unresolved system is $\simeq 0.25$  so the maximum of
 $f_{pb} \, \mathbf{v}_{inn}$ is then $\sim 3.2 \kms$. Larger masses and non-circular orbits will 
  allow slightly higher values, but there is also a reduction by sky projection, so overall   
   the velocity effect of inner binaries is much smaller than we may expect from 
 the familiar $29.8 \kms$ {\newa for a $1 \msun$ binary at $1 \au$ circular orbit }. 
  There is indeed a notable decline in density in Figure~\ref{fig:rpvp} above
  $\dvp \sim 3 \kms$; 
  This corresponds to $\vtilde \sim 3$ for a $ 1 \kau$ outer
 orbit, or $\vtilde \sim 10$ for a $10 \kau$ outer orbit.  So, the tail of triples to large $\vtilde$ 
 is bounded as observed in Figure~\ref{fig:vtilde-trips},  but this bound increases $\propto \sqrt{\rp}$. 
  This also suggests that our original selection cut of $\dvp \le 5 \kms$ above 
    is wide enough to include almost all bound binary and triple systems at $\rp \ge 1.25 \kau$; 
 {\newa none of our simulated triple systems at $\rp \ge 1.25 \kau$ exceeded $\dvp > 5 \kms$. }

	\subsection{Random Samples}
	\label{sec:randoms}
 As a refinement of \citet{Pittordis_2023}, we also include in the fitting a population of ``random" associations:
 this population was generated in PS23 by randomising single star positions by a few degrees (keeping other
 parameters unchanged) then re-running the binary search, and repeating 9 times. 
  Since the number of randoms per bin is relatively small, the statistical uncertainties per bin are rather large.  
 Therefore, we simply create a smooth population of randoms normalised to the mean total per simulation.  
  Since the Galactic rms velocity dispersion is considerably larger than the $5 \kms$ maximum velocity
 difference considered here,  phase-space
 considerations show that the expected population of randoms follows a distribution function given by 
   $dN_r \propto v_p \, r_p \, dv_p \, dr_p$.   
 Changing variables from $v_p$ to $\vtilde$ leads to $dN_r \propto \vtilde \, d\vtilde \, dr_p$.  We adopt this
  distribution scaled to match the measured mean number of randoms per run 
  with $\vtilde \le 7$ in the outer bin $14.1 < r_p < 20 \kau$, which is 37. 

It will be seen below that the randoms make a very small contribution to the final 
 fits, except in the outermost 2 bins. 
 
	\subsection{Flyby simulations} 
	\label{sec:flybys}  
 Similar to PS23, we simulate a population of unbound low-velocity ``flyby" systems; this is assumed 
 to result from unbound pairs which have very similar 3D velocities due to having been born in the
 same open cluster but escaped independently; this means that they retain a ``memory" of their 
 initial orbits and have velocity differences much smaller than random unassociated pairs. 
{\newa An alternative source of such systems is given by \citet{Jiang_2010}, who modelled 
 evolution of wide-binary systems under random encounters with other stars and the Galactic tidal field; 
 they found that binaries wider than $\sim 20 \kau$ have a substantial probability of becoming unbound within the 
  age of the Galaxy, but can remain within separations of a few pc for a long time after unbinding, and occasionally
  approach closer. } 
  
We simulate these as hyperbolic orbits with asymptotic velocities $v_\infty$ uniform and random
 from 0 to $2 \kms$, and model a realistic distribution of impact parameters; we take random
 snapshots of these orbits,  discard systems with
 3D separations above $200 \kau$ and ``observe" the surviving pairs from random viewing directions. 
  Note that in 3D space, these hyperbolic systems must have $\vtilde_{3D} > \sqrt{2}$, but in 2D the projection
 factors fill in the hole at $\vtilde < \sqrt{2}$ giving a smooth distribution. It turns out that the 
 shape of the $\vtilde$ distribution 
  for any single $\rp$ bin is not very different from the triples,  however the flyby population increases 
 steeply with $\rp$, while the triple population decreases with $\rp$ roughly in proportion to the binaries; 
 see below for implications for the fitting.

%%%% FFFFFFFFF  fits  
\section{Data vs model fitting}
\label{sec:fits}  
	
In this section we fit the observed distributions of $\vtilde$ in our DR3 wide-binary sample 
as a mixture of binary, triple, flyby and random systems, where 
 the shape of the $\vtilde$ distribution for each population is fixed by the results of the simulations 
 in Section~\ref{sec:orbits}, while the relative normalisations of each population are adjusted 
 to fit the data.   We repeat this for both Newtonian and MOND gravity models, and several
 eccentricity distributions. 

 We take the bins as defined above in projected separation spanning $1.25 \kau$ to $20 \kau$ in total;  
  this range is split into 8 logarithmic bins 
 with each bin upper edge $\sqrt{2} \times $ the lower edge, so the first bin spans $1.25$ to $1.77 \kau$ and
 the eighth bin is $14.1$ to $20 \kau$; this places the first two bins in the quasi-Newtonian regime where
 MOND effects should be small, bins 3 -- 5 are in a transition regime, while the final three bins are well into the MOND regime with internal accelerations $\la a_0$. 
 We then fit simultaneously to all eight observed $\vtilde$ histograms, using 9, 10 or 11 adjustable parameters: 
 below we first describe our ``baseline" 10-parameter model fit in some detail in the next subsection, then 
  explore some variations in the base fits in the subsequent subsection. 

\subsection{Base model fits} 
\label{sec:fitbase} 

 Our base model uses 10 adjustable parameters for normalization of the model distributions:  
  the first 8 parameters $n_{bt, j}$ for $1 \le j \le 8$ are simply
 the total number of model binary+triple systems in the $j$th bin, 
 the 9th parameter $\ftrip$ is the fraction of triples among (binaries + triples), kept equal across all bins;  
 and the 10th parameter $n_{fly}$ is one overall normalisation for the flyby population 
  (with the relative numbers per bin fixed as in the simulation).  
 The random population is included in the model fitted, but fixed as in the previous section.  
 Then, for each bin the model populations are normalised such that the bin $j$  contains
  $(1 - \ftrip) \, n_{bt, j}$  binaries, $\ftrip \, n_{bt, j}$ triples, the flyby model is renormalised
  so there are $n_{fly}$ flybys in the $7-10 \kau$ bin (this bin arbitrarily defined),   
   and all other bins are rescaled by the same factor 
 so relative number of flybys per bin remains the same as the simulation;  
 and the model randoms are fixed as above with pre-defined normalisation. 
  These four populations are then added to give a total model $\vtilde$ histogram for each of the 8 bins. 
 (Note that the use of a single normalisation parameter for the flyby population across all bins is helpful here:  
 in our previous paper \citetalias{Pittordis_2023},  the four $\rp$ bins were fitted independently 
 which led to a substantial degeneracy between flyby and triple populations.  With the simultaneous
 fit as here, the $\rp$ distribution of the flybys is fixed, which breaks this degeneracy.)   

 We then fit a maximum-log-likelihood fit to the observed 
 histograms of $\vtilde$,  varying these 10 parameters to maximise the summed log-likelihood  of all 8 bins combined, using the Python {\tt lmfit} fit package and a Nelder-Mead simplex search; log-likelihoods are calculated using the data and Poisson statistics from the total model distribution as above (this is formally better than a $\chi^2$ minimisation, due to small-number statistics in the tails at  high $\vtilde$).  

\begin{figure*}
\includegraphics[width=15cm]{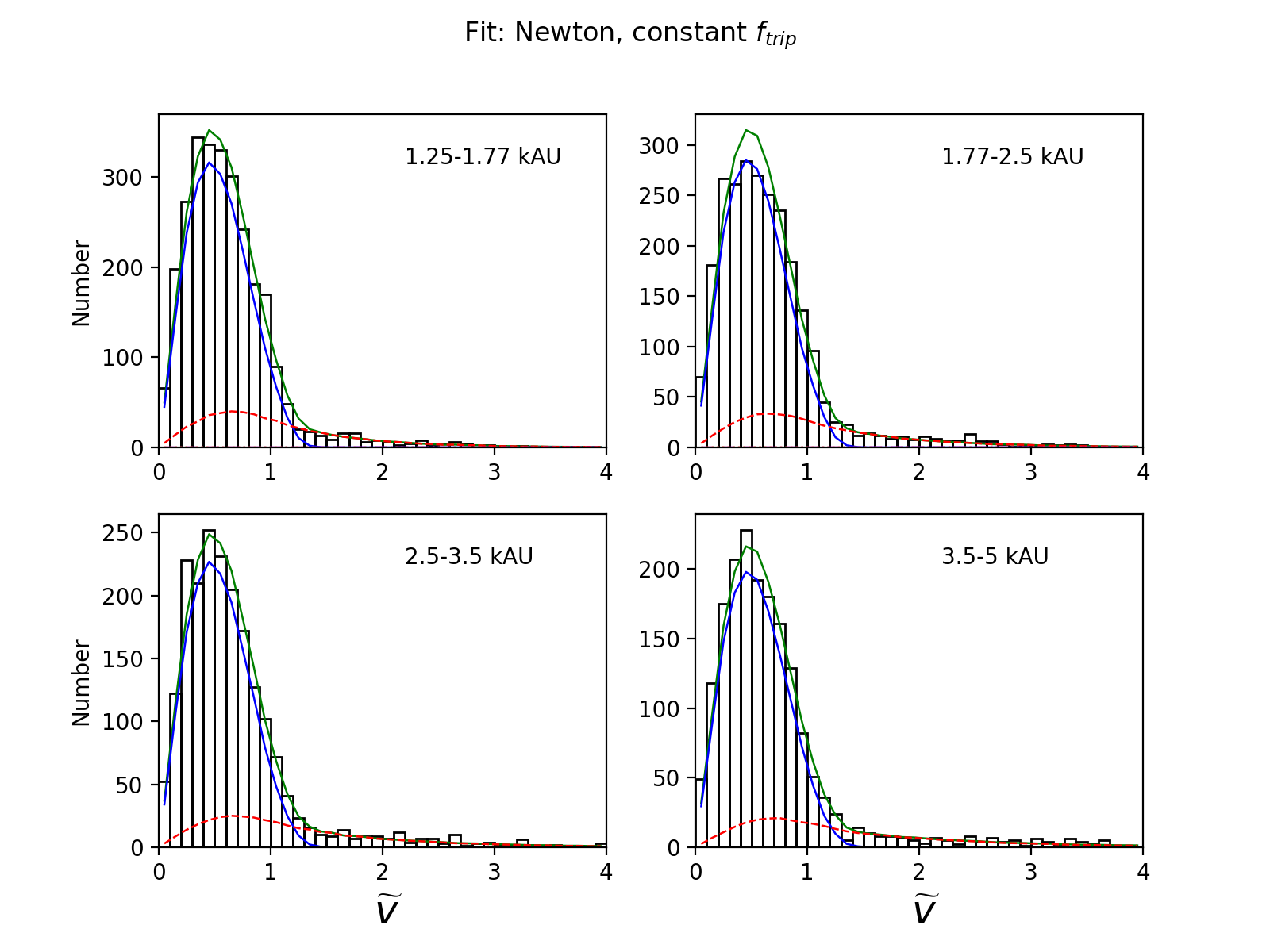} 

\includegraphics[width=15cm]{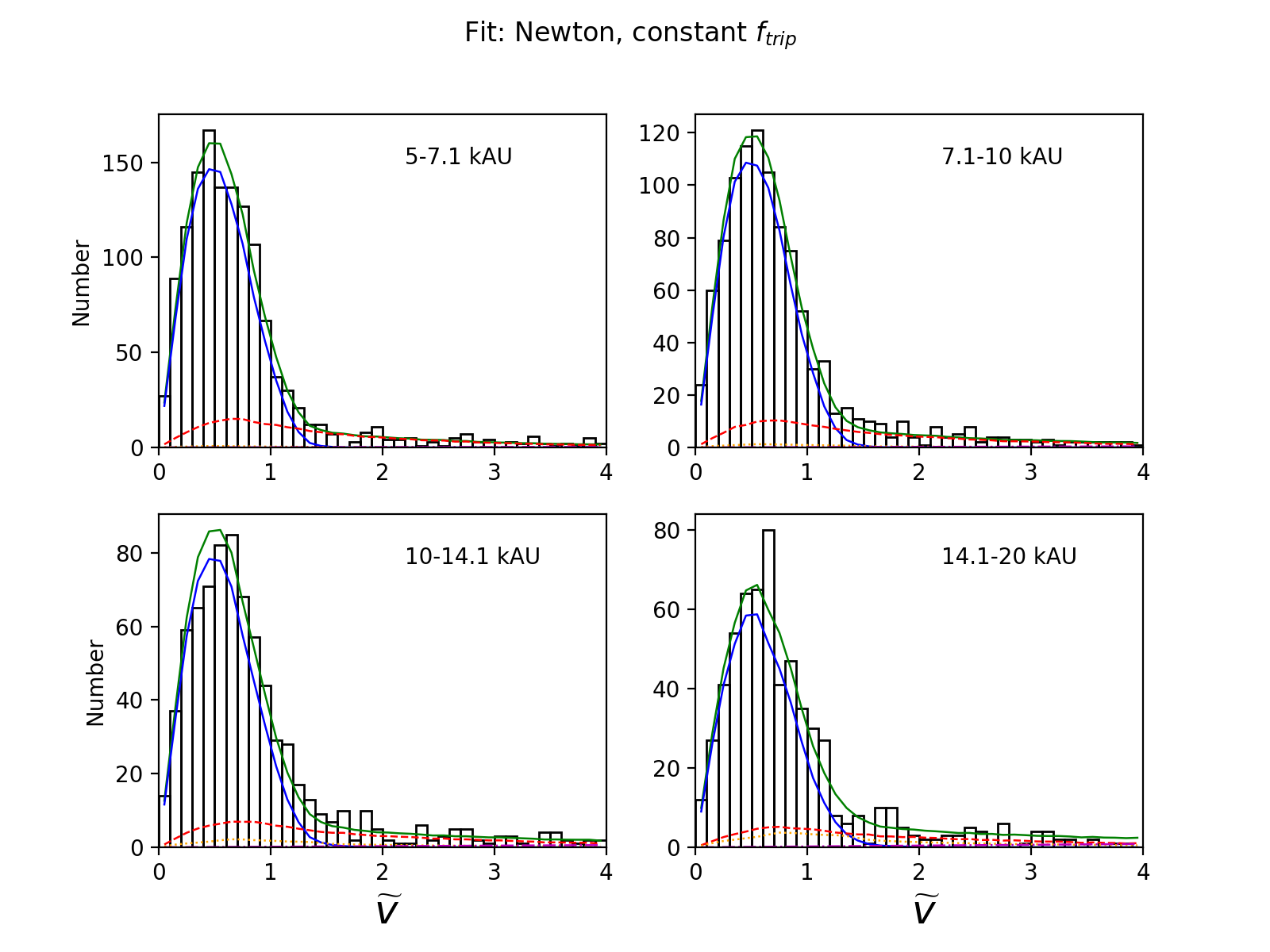}

\caption{Fits of observed $\vtilde$ histograms for the baseline Newtonian models.  The 8 panels show different $r_p$ bins as in the legend.  Data is the black histogram.  Lines show the total model fit (solid green, top), and subcomponents: binaries (solid blue, lower), triples (dashed red), flybys (dotted orange) and randoms (dot-dash magenta).  The randoms are very low values except in the last two bins,  \label{fig:basefit_newt} 
} 
\end{figure*}

\begin{figure*}
\includegraphics[width=15cm]{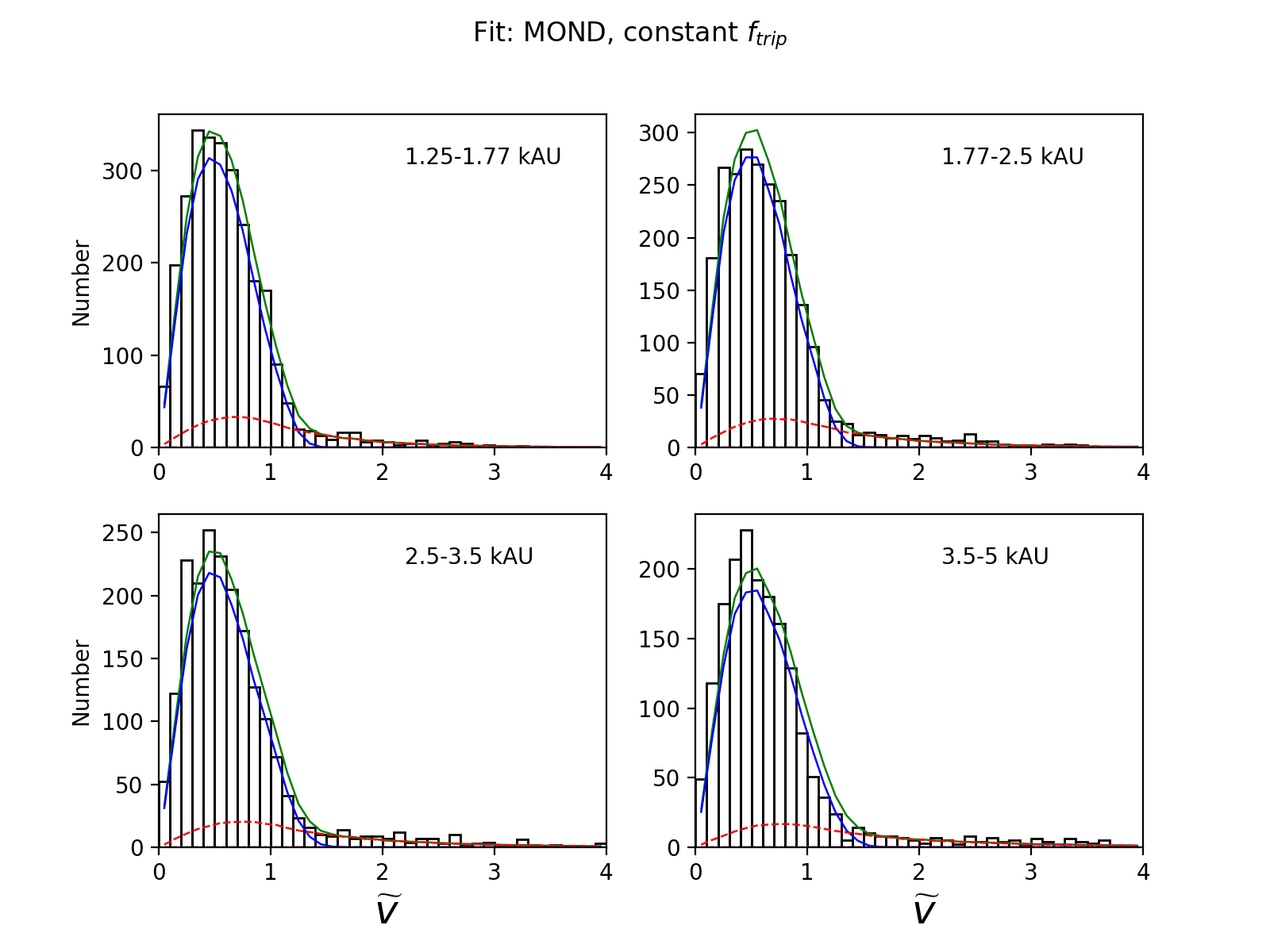} 

\includegraphics[width=15cm]{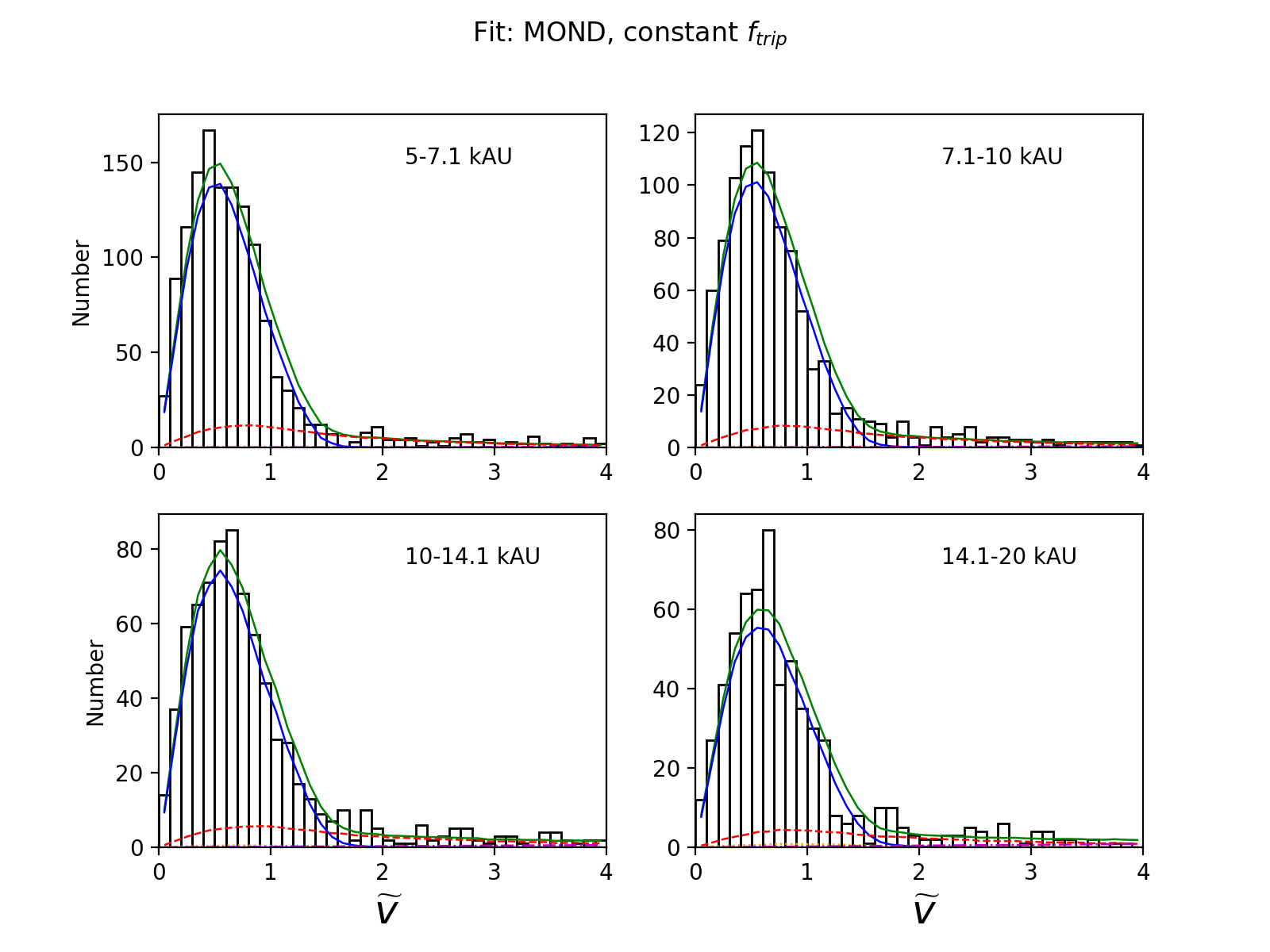}

\caption{Same as Figure~\ref{fig:basefit_newt}, but fitted with MOND model binary and triple
   $\vtilde$ distributions. 
 \label{fig:basefit_mond} 
} 
\end{figure*}

We then run the fit as above for both the Newtonian and MOND binary+triple orbit models: 
  flyby and random models are in common,  but all the parameters are independent between the two fits.  
The results of our baseline fits are shown for the Newtonian model in Figure~\ref{fig:basefit_newt}, for
 MOND in Figure~\ref{fig:basefit_mond},  and best-fit parameters listed in Table~\ref{tab:basefit}. 

%% para updated for 2025-05-31 model 
The most notable feature is that the best-fit Newtonian model produces a  dramatically better $\chi^2$
 than the best-fit MOND model, respectively 424 for  Newton vs 704 for MOND, i.e. a difference $\Delta \chi^2 = 280$ which is formally extremely significant.  The expected $\chi^2$ given the $8 \times 40$ histogram data values 
 fitted is 310, so both fits are formally unacceptable;  however the Newtonian model has a $\chi^2$ per degree of freedom of 1.37 which is not unusual for real-world data. If we divide both $\chi^2$ values by this factor, then the 
 $\Delta \chi^2$ for MOND-Newton becomes 205 which is still very large;  it appears that the Newtonian
 fit is preferred at very high significance. 

On inspection of the Figures, both models produce a reasonable fit at small $\vtilde \la 0.5$ and in 
 the extended tail $\vtilde > 1.5$ which is triple-dominated; the main deviation of the MOND fits
 is that the MOND models systematically over-predict the data in the key region $0.8 \le \vtilde \le 1.2$, due to the predicted ``rightward stretch" of the binary population predicted by MOND: 
 this stretch therefore appears {\bf not to be present} 
 in the data.   The Newtonian fits are much closer to the data in this key $\vtilde$ region. 

\begin{figure} 
\begin{center} 
\includegraphics[width=\linewidth]{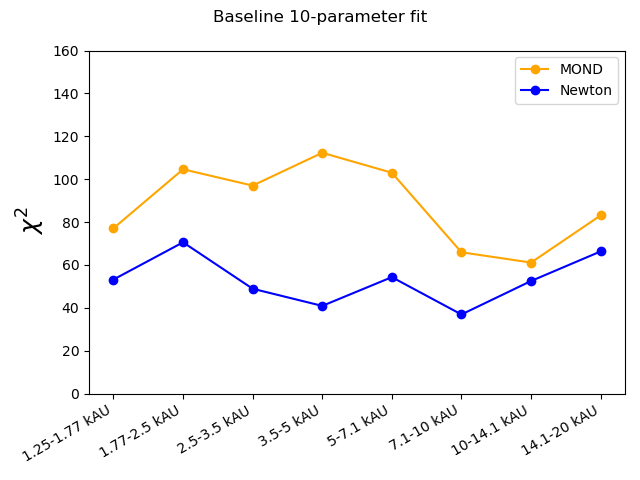}    % updated post-referee 
\caption{ The fit $\chi^2$ values, for the baseline fit, 
  for individual $\rp$ bins (x-axis) for Newtonian and MOND models 
\label{fig:chisq_rp} 
} 
\end{center}
\end{figure}

We show the individual $\chi^2$ values per bin for both models in Figure~\ref{fig:chisq_rp}; 
 this shows that all the individual bins prefer Newtonian, with the strongest preference in the $3.5 - 5 \kau$ bin. This
 is slightly counter-intuitive because the MOND excess velocity grows with separation; however there are
 a larger number of systems in this bin, and the triple tail is less prominent, so this may well explain the result. 
 The final two bins suffer somewhat from small-number statistics in the key region $0.8 \le \vtilde \le 1.4$.  

{\newa 
Another point to note is that the $\Delta \chi^2$ (MOND-Newton) does decrease at small $\rp$, but 
 does not reach zero.  A probable reason is as follows: in our first bin $1.25 \le \rp \le 1.77 \kau$, 
 model binaries have a median current 3D separation of $1.7 \kau$, and the MOND effect at this 
 separation is well below 1 percent excess over Newtonian, essentially negligible. 
 However, the velocity depends on the MOND effect averaged over a full orbit; for this $\rp$ bin 
 25\% (10\%) of the model binaries have an apocenter separation above $3.2\ (5.2) \kau$ respectively, where 
 the MOND effect reaches 4 (15) percent.  Thus the orbit-averaged MOND effect is on average
  larger than the instantaneous value.  } 

\begin{table*} 
\caption{ Fit parameters for baseline fit, Newton and MOND models  \label{tab:basefit} 
} 
\begin{center} 
%%%    Here is a 1+3+3-column version,  commented out  ....  actual 1+2+2 column  below 
%\begin{tabular}{lccc|ccc } 
% & \multicolumn{3}{c}{Newton} & \multicolumn{3}{c}{MOND} \\ 
%\cline{2-4} \cline{5-7} 
%  here this is an ell-vert  to put line at RHS of entry 
% & \multicolumn{3}{l|}{ $ \ftrip = 0.170 \ \ n_{fly} = 18.0 $ }  &  \multicolumn{3}{l}{ \quad $\ftrip = 0.143 \ \ n_{fly} = xxx$ } \\
% $r_P$ bin   &  $n_{bt}$ & $\chi^2$ &  xxx  &  $n_{bt} $  & $ \chi^2$ & xxx \\ 
%\hline 
% $1.25 - 1.77 \kau$ &  xxx &  xxx & xxx &  xxx & xxx & xxx \\ 
% $1.77 - 2.5 \kau$ &  xxx &  xxx & xxx &  xxx & xxx & xxx \\ 
% $2.5 - 3.5 \kau$ &  xxx &  xxx & xxx &  xxx & xxx & xxx \\ 
% $3.5 - 5.0 \kau$ &  xxx &  xxx & xxx &  xxx & xxx & xxx \\ 
% $5.0 - 7.1 \kau$ &  xxx &  xxx & xxx &  xxx & xxx & xxx \\ 
% $7.1 - 10 \kau$ &  xxx &  xxx & xxx &  xxx & xxx & xxx \\ 
% $10 - 14.1 \kau$ &  xxx &  xxx & xxx &  xxx & xxx & xxx \\ 
% $14.1 - 20 \kau$ &  xxx &  xxx & xxx &  xxx & xxx & xxx \\ 
% \end{tabular} 
% 
%% Actual table : 1+2+2 column  - now updated with 04apr model fits. 
\begin{tabular}{lcc|cc } 
 & \multicolumn{2}{c}{Newton} & \multicolumn{2}{c}{MOND} \\ 
\cline{2-3} \cline{4-5} 
%  here this is an ell-vert  to put vertical line at RHS of entry 
%% Following 12 lines are old data from v1 for comparison 
% & \multicolumn{2}{l|}{ $ \ftrip = 0.170 \ \ n_{fly} = 18.0 $ }  &  \multicolumn{2}{l}{ \quad $\ftrip = 0.143 \ \ n_{fly} = 6.6 $ } \\
% $r_P$ bin   &  $n_{bt}$ & $\chi^2$ &   $n_{bt} $  & $ \chi^2$  \\ 
%\hline 
% $1.25 - 1.77 \kau$ &  2723 &  80.6       & 2724 &  148.4 \\ 
% $1.77 - 2.5 \kau$ &  2455 &  85.0       & 2456 &  163.5 \\ 
% $2.5 - 3.5 \kau$ &  1964 &  71.7        & 1966 & 170.6 \\
% $3.5 - 5.0 \kau$ &  1728 &  59.1             & 1731 &  189.5  \\
% $5.0 - 7.1 \kau$ &  1305 &  62.4         & 1310 & 163.5 \\ 
% $7.1 - 10 \kau$ &  989 &   39.4          & 1001 & 103.6 \\ 
% $10 - 14.1 \kau$ &  719 &  49.1        & 740 & 77.4 \\ 
 % $14.1 - 20 \kau$ &  547 &  64.0        & 581 & 91.8 \\ 

%% This table now updated with 2025-05-31 model run. 
& \multicolumn{2}{l|}{ $ \ftrip = 0.185 \ \ n_{fly} = 17.9 $ }  &  \multicolumn{2}{l}{ \quad $\ftrip = 0.155 \ \ n_{fly} = 4.1 $ } \\
 $r_P$ bin   &  $n_{bt}$ & $\chi^2$ &   $n_{bt} $  & $ \chi^2$  \\ 
\hline 
 $1.25 - 1.77 \kau$ &  2723 &  53.2    & 2724 &  77.1 \\ 
 $1.77 - 2.5 \kau$ &  2455 &  70.6     & 2456 &  104.6 \\ 
 $2.5 - 3.5 \kau$ &  1964 &  48.9       & 1966 & 97.1 \\
 $3.5 - 5.0 \kau$ &  1728 &  40.9       & 1731 &  112.3  \\
 $5.0 - 7.1 \kau$ &  1305 &  54.3       & 1312 & 103.0 \\ 
 $7.1 - 10 \kau$  &  990  & 36.9         & 1004 & 65.9\\ 
 $10 - 14.1 \kau$ &  720 &  52.5        & 747 &  61.1 \\ 
 $14.1 - 20 \kau$ &  548 &  66.3        & 590 &  83.0 \\ 
\end{tabular} 
\end{center} 
\end{table*} 

 \subsection{Variations on the base model} 
\label{sec:fitvars} 
The results above appear highly significant, but with the caveat that they are limited to specific model shapes
 of the binary and triple populations {\newa and assumptions about unbound (flyby + random) populations. } 
To check for sensitivity to various assumptions, we have explored a number of variations on the base
 model as follows: 

{\bf Scale-dependent $\ftrip$:}  Here we add an 11th free parameter, by allowing $\ftrip$ to be a linear function 
 of $\log \rp$; specifically we use $ \ftrip = c_0 + c_1 \, j $ where $j$ is the bin number, and fit for $c_0, c_1$ 
 instead of $\ftrip $.  The result is the best-fit values are $c_0 = 0.173$, $c_1 = 0.0058$, so the variation in $\ftrip $ is relatively modest, with the 8th bin having $\ftrip = 0.214$;  
  the Newtonian $\chi^2$ is 420, a reduction of only 4 from the base model, 
  and the $\Delta \chi^2$ (MOND-Newton) is 284. 

{\bf Zero flybys, default randoms: }  Here we use a 9-parameter fit by fixing the flyby parameter $n_{fly}$ to zero. 
 The result is that $ \ftrip $ increases to 0.193, the Newtonian $\chi^2$ worsens to 443, and the 
 $\Delta \chi^2$ (MOND-Newton) is 264. 

{\bf Zero flybys, floating randoms: }  This is a 10-parameter fit with the flyby parameter $n_{fly}$ again fixed to zero, but allowing the normalisation of the randoms to float. The Newtonian results give random normalisation $0.75$, 
 $\ftrip = 0.195$ and $\chi^2 = 446$.   The $\Delta \chi^2$ (MOND-Newton) is 264. 

{\bf Double randoms: } This is the same as the 10-parameter base fit except that the population of randoms
 is doubled from the default value.  The result is that $\ftrip  = 0.182$, the Newtonian $\chi^2$ is 435, 
 and the $\Delta \chi^2$ (MOND-Newton) is 269 

{\bf Base model, no perspective effects: } This uses the same 10-parameter base fit, but the model 
 binary and triple populations do not include perspective effects from Section~\ref{sec:perspec}. 
 The result is $ \ftrip = 0.186$, the Newtonian $\chi^2$ is 423, and the $\Delta \chi^2$ (MOND-Newton) is 273.  . 

In summary, these variations on the base fit show that the preference for Newtonian gravity over MOND is not sensitive to variations in the random and flyby populations:  the main feature of the fits is that
   the tail population at $\vtilde > 1.5$ is
  dominated by triples, with the randoms and flybys a minor contribution except in the widest $\rp$ bin of 
 $14.1 - 20 \kau$.  
 This property is well constrained, because the randoms and flybys cannot over-predict the tail in the latter bin; then   
  their abundance falls off steeply in smaller bins due to their model $\rp$ distributions, while the triples become more numerous towards smaller bins pro-rata with the binaries.   This requires that the high $\vtilde$ tail in most $\rp$ bins is  dominated by triples.   
 Therefore,  the observed amplitude of the tail constrains $\ftrip$ rather well.  Given this, the models
 give a smooth extrapolation of the triples downwards below $\vtilde < 1$, so unless our triple model is
  for some reason substantially too high at $\vtilde \sim 1$, it seems hard to avoid the MOND models 
  overshooting the data in this region.

\subsection{Implications for triple fraction} 
\label{sec:ftrip} 
The Newtonian fits above give a triple fraction of $\ftrip = 0.185$ for the base model, with fairly small
 variations in the alternative fit models.  The MOND model fits are systematically lower than this by $\approx 0.03$, 
  in the direction expected due to the more extended underlying distributions in MOND. 

%% numbers updated from 2025-05-31 model run 

Note that this $\ftrip$ is the fit {\em after} the simulated data-cuts are applied to the models.  We can infer the fraction of triples pre-cut as follows. On average 38.7 percent of our simulated triples survive the data-cuts. The inferred pre-cut triple fraction is somewhat smaller than $0.185 / 0.387$ or 48 percent,  since the denominator also includes triples. Assuming that nearly all pure binaries survive the data cuts,  a population ratio of 18.5 triples to 81.5 binaries post-cuts would imply 48 triples per 81.5 binaries before the cuts,  hence an original triple fraction of $48/(48+81.5)$ hence 37 percent; if the cuts reject a few binaries, this would be lowered slightly.  
 This is significantly lower than the best-fit value $f_{CB} = 0.63$ in the nominal fit of \citet{Banik_2024}, and perhaps more realistic compared  with observational estimates.  We note that \citet{Banik_2024} also fitted the gravity model with $f_{CB}$ constrained to a lower value of 0.30 (last row of their Table 3), and still found a strong preference for Newtonian gravity  consistent with our results. 

  The 37 percent appears reasonable: e.g.  in the WB formation model suggested by \citet{Kouwenhoven_2010}, wide-binaries form from capture of independently-born close systems during dissolution of an open cluster. 
{\newa  
 A detailed study of multiplicity for solar-neighbourhood F/G dwarfs is given by \citet{Tokovinin_2014}, 
  with the result that systems with F/G primaries are
54\% single, 33\% binary, 8\% triple, 4\% quadruple, 1\% higher order. 
 % Very similar results were reported by Raghavan et al (2010) arXiv:1007.0414.
Using the Kouwenhoven et al estimate that 15\% of binaries are "wide" splits the binary contribution into 
 28\% close binary, 5\% wide binary,  Assuming all the wide-binary systems originated from capture of two primordial single systems and that the remaining multiples are primordial, then the fraction of primordial single vs multiple systems would be 64/105 = 61\%.   
%% tweaked this a bit. 
We do not have comprehensive statistics for the lower-mass spectral types, K and early-M, which make up slightly over half our sample, but it is known that multiplicity is positively correlated with primary mass, so a primordial  single fraction nearer 70\% seems reasonable averaged over our mass distribution.   

If we also assume that wide systems form from capture of two independent close systems, 
 this would imply that wide systems are 49\% pure binaries, and 51\% triples or higher-order multiples.  This 51\% of triples plus higher-order multiples is somewhat higher than our estimated 37\%, but is not seriously discrepant.  } 
\vfill  

\subsection{Caveats and Limitations} 	
\label{sec:caveats} 
 The main limitations of our study is that we have only a single MOND model, and
 the parameter space of triple models is rather limited, though it is consistent
 with recent data from \citet{Offner_2023} and \citet{Hwang_2022}. 
  The related wide-binary study by \citet{Banik_2024} 
 used a more flexible parameter space of triple models, and reached the same preference
 for Newtonian over MOND gravity. 

 Another limitation is that we have not yet included quadruple (2+2) systems  
 in our modelling; this is left for future work. 

{\newa Here we have assumed that the estimated errors in \GAIA proper motions are realistic, 
  hence adding simulated magnitude-dependent errors to our modelled values.  If the \GAIA
 proper motion errors are systematically underestimated, or there is a strong non-Gaussian tail, 
 this would tend to shift observed $\vtilde$ values systematically upwards, with the
 relative effect increasing with $\rp$.  This  would tend to produce
 a false-positive MOND signal in the data, but this is in the opposite direction to our
  conclusion so is unlikely to be a serious problem.  }  
	
 {\newa In an earlier verson of this paper, the $\Delta \chi^2$ values (MOND-Newton) were significantly
 higher at $\sim 500$ compared to $\sim 280$ above; both Newtonian and MOND $\chi^2$ values
 are lower, but the MOND values lowered from $\sim 1000$ to $700$ while Newtonian 
 values lowered from $\sim 500$ to $420$.    The main reason for this shift is the change
 from a flat distribution in $\log a_{out}$ (i.e. $ p(a_{out}) \propto a_{out}^{-1}$  to a steeper power-law 
$\propto a_{out}^{-1.6}$.  We find that this shifts the pure-binary distributions of $\vtilde$ by approximately a multiplicative shift of approx $0.96 \times$. The reason is as follows: in 3D the $\vtilde_{3D}$ in Newtonian gravity is simply 
$\sqrt{2 - r/a}$; we observe in bins of $\rp$, while $a$ is unknown.  At a given $r$, systems have a range of $a$ 
 from $a \sim r/2$ to $a \gg r$. 
 Those with  $a > r$ have $\vtilde_{3D} > 1$ and those with $a < r$ have $\vtilde_{3D} < 1$.  
 Changing to a more steeply declining $a_{out}$ distribution leads to more of the latter and fewer of the former  at a given $r_p$, hence the systematic shift.  The $4\%$ downward shift (in both Newtonian and MOND) 
  is about 1/4 of the large-separation MOND effect, so the new MOND fits are  less bad than previously.   

}  % end new para 
 
	%% new subsection for discussion of Hernandez et al... 
\subsection{Comparison with previous studies} 
Several earlier studies of wide binaries in \GAIA EDR3/DR3 by \citet{Hernandez_2022}, \citet{Hernandez_2023},
 \citet{Hernandez_VNA_2024} 
 \citet{Chae_2023}, 
 \citet{Chae_2024} have shown a strong preference for MOND over Newtonian gravity, thus 
 directly opposite to our conclusion above; a review is given by \citet{Hernandez_CA_2024}.  

 Most of these analyses used fairly stringent cuts to remove triple systems, but then the surviving
 samples were analysed assuming zero residual contamination from triples.   We argue here that this
  assumption is unlikely to be realistic, since triples with inner separations $\sim 5 - 50 \au$ and dissimilar
 masses are not removable by any current cut: 
 their orbit periods are too long to be rejected by a {\tt ruwe} cut, their angular separations are too close
  to be resolvable, and mass ratios $q \la 0.75$  will  not be rejected by a Lobster-type cut;  
 so systems like this will  inevitably contaminate any present-day sample, at least until dedicated followup data 
 is taken.  Our modelling above indicates that $38.7$ percent of our simulated triple systems survived all our
 cuts;  this is clearly not a small fraction, 
 therefore we believe that accounting for  residual triples in the  fit procedure is more realistic. 

Residual triples have $\vtilde$ values systematically larger than binaries, and for fixed inner orbits
 this shift increases with the outer orbit $\rp$ due to the $1/\sqrt{\rp}$ term in the denominator of $\vtilde$; 
 so unaccounted residual triples tend to produce a MOND-like signal. 

\citet{Chae_2024} does include model triples in the fits: however one issue is that his $\ftrip$ is 
 normalised to match the median $\vtilde$ in a small-separation bin, then held fixed in all other bins.
 This is questionable since the median $\vtilde$ at small separations is not very sensitive to triples; another issue
 is that Chae's sample is derived from \citet{Badry_2021} and thus contains a cutoff 
 $\dvp \le 2.1 \kms (\rp/1 \kau)^{-0.5}$; converting to $\vtilde$ gives $\vtilde \le 2.24 / \sqrt{M_{tot}}$. 
 Therefore the extended tail at $\vtilde > 2$ is almost entirely removed from the Chae sample which 
 greatly reduces the leverage on $\ftrip$.  It is notable that the right columns of 
  Figs. 15-18 of \citet{Chae_2024} all show a systematic
 excess of data above models in the range $1.5 \le \vtilde \le 2$,  perhaps indicating a problem in the triple modelling.

Further progress in the WB gravity test will likely depend on improved constraints and understanding of 
   the triple population: in a first step in this direction, \citet{Manchanda_2023} have shown 
 that combining acceleration signals from the 10-year \GAIA data plus followup speckle imaging and/or
 coronagraphic imaging will be able to detect almost all triples with a main-sequence third star 
 at any separation.  (Triples with a brown dwarf third object are much more challenging).

 %%%% CCCCC conclusions 
	\section{Conclusions}
	\label{sec:conc} 
 We have used a sample of wide binaries selected from \GAIA DR3 to perform a test of 
 a specific MOND model (with external field effect) against GR/Newtonian gravity.  This includes a number of refinements  from our earlier study in \citetalias{Pittordis_2023}, including several new selection cuts to 
 discriminate against hierarchical triples; a wider range of $\rp$ in the analysis; updates to 
 the triple model with more realistic mass and semi-major axis distributions; velocity-averaging
 for short-period inner orbits; and an improved fitting procedure which simultaneously fits
 multiple $\rp$ bins with a flexible triple fraction $\ftrip $ plus unbound flyby and random-projection populations.

 With our baseline triple model, the results show a rather strong preference for Newtonian gravity over the MOND model in question: this preference is robust against several variations on the baseline fit. 
 The implied residual triple fraction is $ \ftrip \simeq 18.5 $ percent  (after the cuts to reject most triples). 
 The implied triple fraction {\em before} our triple-rejection cuts with {\tt ruwe}, {\tt ipd\_frac\_multi\_peak} , faint companion and Lobster cuts is approximately 37 percent, lower than the 63 percent inferred 
 by the fit of \citet{Banik_2024}; this appears reasonable based on the formation model of 
\cite{Kouwenhoven_2010}. 

We cannot claim that MOND is fully ruled out as yet: while our triple model 
 is based on observational estimates,    there remains a possibility that an alternative 
 triple model may result in a lower triple fraction and/or a less poor fit to MOND. However, this seems
 unlikely for two reasons: first,   
 we note that our triple model went through a number of upgrades during the course of this work: 
 all our earlier iterations also preferred Newtonian gravity over MOND with fairly high $\Delta \chi^2$; 
 so the triple model has been refined to match recent parameter estimates in the literature, 
 but has not been fine-tuned to prefer  Newtonian gravity.  
  Second, it may be observed in Figure~\ref{fig:basefit_mond} for the MOND fits, the contributions from
  {\em binaries alone} 
 are already tending high compared to the data  at $0.8 \le \vtilde \le 1.2$ in the  upper few $\rp$ bins. 
 Since the triple and other contributions cannot be negative, and in practice the triples must be significantly positive in this region in order to fit the tail at $\vtilde > 1.5$,  reasonable modifications of the triple distribution will not remove this mismatch. 

 There are good prospects for improving this test: the upcoming \GAIA DR4 release 
 in 2026 will provide dramatically improved proper-motion precision, allowing expanding the 
  sample to $G \sim 18$ and distance $\sim 400 \pc$ for improved statistics.   The full-epoch astrometric
 data will also allow fitting for accelerations from long-period inner orbits, though the full power
  of this test will need to wait for the 10-year \GAIA DR5 around 2030, as modelled in \citet{Manchanda_2023}. 
In the medium term future, combining the improvements from \GAIA DR4/DR5 with
  additional followup to directly detect most triples and improve the constraints 
 on the triple/quadruple population, there are excellent prospects for the wide-binary gravity test to 
 become decisive. 

	\section*{Acknowledgements} 
	We thank Indranil Banik and Kyu-Hyun Chae for many helpful discussions, and  we thank
 two anonymous referees for comments which improved the paper. 
 We thank the great effort of the \GAIA DPAC team for the exceptional quality of the data. 
 We have used the Python packages Numpy, Scipy, Matplotlib, Astropy, Lmfit, and Scikit-learn.  

	\section*{Data availability}
	The original data used in this paper is publicly available.
	The \GAIA data can be retrieved through the \GAIA archive
	(\url{https://gea.esac.esa.int/archive}).

	% put this back for final arXiv submission. 
	%This is an author-produced, non-copy-edited version of the paper as
	%accepted by MNRAS. The version of record is available at 
	% DOI:10.1093/mnras/abc9999  . 
	%\fi

	%%%%%%%%%% RRRRRRR  
	

	\clearpage    % put this back if trailing figures... 
	\bibliographystyle{aa_url} % style aa.bst
	\bibliography{Refs_PSS25} % your references Yourfile.bib
	
	%%%%%%%%%%%%%%% %% APPENDICES %%%%%%%%%%%%%%%%%%%%%
	\appendix
\section{SQL query} 
    \label{app:sql} 
 The following is the SQL query used to select the initial single-star dataset from the \GAIA archive. 

	\begin{verbatim}
		SELECT
			gaiadr3.gaia_source.*,
			gaiadr3.astrophysical_parameters.radius_flame,
			gaiadr3.astrophysical_parameters.radius_flame_lower,
			gaiadr3.astrophysical_parameters.radius_flame_upper,
			gaiadr3.astrophysical_parameters.lum_flame,
			gaiadr3.astrophysical_parameters.lum_flame_lower,
			gaiadr3.astrophysical_parameters.lum_flame_upper,
			gaiadr3.astrophysical_parameters.mass_flame,
			gaiadr3.astrophysical_parameters.mass_flame_lower,
			gaiadr3.astrophysical_parameters.mass_flame_upper,
			gaiadr3.astrophysical_parameters.gravredshift_flame,
			gaiadr3.astrophysical_parameters.gravredshift_flame_lower,
			gaiadr3.astrophysical_parameters.gravredshift_flame_upper,
			gaiadr3.astrophysical_parameters.flags_flame
		FROM
			gaiadr3.gaia_source
		LEFT OUTER JOIN
			gaiadr3.astrophysical_parameters
		ON
			gaiadr3.astrophysical_parameters.source_id 
			= gaiadr3.gaia_source.source_id	
		WHERE
			gaiadr3.gaia_source.parallax >= (10.0/3.0) 
		AND
			gaiadr3.gaia_source.phot_g_mean_mag <=17
		AND
			(ABS(gaiadr3.gaia_source.b) >= 15)
		ORDER BY 
			gaiadr3.gaia_source.dec ASC
	\end{verbatim}
	\label{sec:DR3_FLAMES_QUERY}
		
	\label{lastpage}   % put before end doc.	

%	----------------------------------------------------------------
		
\end{document}